%% file: main.tex
\documentclass[sigconf, nonacm]{acmart}
\usepackage{url}
\usepackage{comment}
\usepackage{graphicx}
\usepackage{subcaption}
\usepackage{balance}  % for  \balance command ON LAST PAGE  (only there!)
\usepackage{framed}
\usepackage{enumitem}
\usepackage{multirow}
\usepackage{xcolor}
\usepackage{textcomp}
\usepackage{wrapfig}
\usepackage{amsmath}

\newcommand\vldbdoi{XX.XX/XXX.XX}
\newcommand\vldbpages{XXX-XXX}
\newcommand\vldbvolume{14}
\newcommand\vldbissue{1}
\newcommand\vldbyear{2020}
\newcommand\vldbauthors{\authors}
\newcommand\vldbtitle{\shorttitle} 
\newcommand\vldbavailabilityurl{http://vldb.org/pvldb/format_vol14.html}
\newcommand\vldbpagestyle{plain} 

\definecolor{brightgreen}{rgb}{0.0, 0.5, 0.0}
\definecolor{ao(english)}{rgb}{0.0, 0.5, 0.0}
\definecolor{amber(sae/ece)}{rgb}{1.0, 0.49, 0.0}
\definecolor{darkorange}{rgb}{1.0, 0.55, 0.0}
\definecolor{darkmagenta}{rgb}{0.55, 0.0, 0.55}

\newcommand\sys[1][]{\textit{BI-REC}}

\begin{document}
	\title{BI-REC: Guided Data Analysis for \\ Conversational Business Intelligence}
	\author{Venkata Vamsikrishna Meduri$^{1*}$, Abdul Quamar$^2$, Chuan Lei$^2$, Vasilis Efthymiou$^{3*}$, Fatma {\"O}zcan$^{4*}$}
% 	\authornote{Work done while at IBM Research}
	\affiliation{%
		\institution{$^{1}$ Arizona State University, $^{2}$IBM Research - Almaden, $^3$ICS-Forth, $^4$Google}
		\streetaddress{}
		\city{}
		\state{}
		\postcode{}
	}
	\email{ vmeduri@asu.edu,ahquamar@us.ibm.com, chuan.lei@ibm.com,vefthym@ics.forth.gr,fozcan@google.com
}
\thanks{*Work done while at IBM Research.}

\begin{abstract}
Conversational interfaces to Business Intelligence (BI) applications enable data analysis using a natural language dialog in small incremental steps. To truly unleash the power of conversational BI to democratize access to data, a system needs to provide effective and continuous support for data analysis. In this paper, we propose \sys, a conversational recommendation system for BI applications to help users accomplish their data analysis tasks.

We define the space of data analysis in terms of BI patterns, augmented with rich semantic information extracted from the OLAP cube definition, and use graph embeddings learned using GraphSAGE to create a compact representation of the analysis state. We propose a two-step approach to explore the search space for useful BI pattern recommendations.
In the first step, we train a multi-class classifier using prior query logs to predict the next high-level actions in terms of a BI operation (e.g., {\em Drill-Down} or {\em Roll-up}) and a measure that the user is interested in. In the second step, the high-level actions are further refined into actual BI pattern recommendations using collaborative filtering. This two-step approach allows us to not only divide and conquer the huge search space, but also requires less training data. 
Our experimental evaluation shows that \sys~achieves an accuracy of 83\% for BI pattern recommendations and up to 2$\times$ speedup in latency of prediction compared to a state-of-the-art baseline.
Our user study further shows that \sys~provides recommendations with a precision@3 of 91.90\% across several different analysis tasks.

\end{abstract}

\maketitle
	
\pagestyle{\vldbpagestyle}
\begingroup\small\noindent\raggedright\textbf{PVLDB Reference Format:}\\
\vldbauthors. \vldbtitle. PVLDB, \vldbvolume(\vldbissue): \vldbpages, \vldbyear.\\
\href{https://doi.org/\vldbdoi}{doi:\vldbdoi}
\endgroup
\begingroup
\renewcommand\thefootnote{}\footnote{\noindent
	This work is licensed under the Creative Commons BY-NC-ND 4.0 International License. Visit \url{https://creativecommons.org/licenses/by-nc-nd/4.0/} to view a copy of this license. For any use beyond those covered by this license, obtain permission by emailing \href{mailto:info@vldb.org}{info@vldb.org}. Copyright is held by the owner/author(s). Publication rights licensed to the VLDB Endowment. \\
	\raggedright Proceedings of the VLDB Endowment, Vol. \vldbvolume, No. \vldbissue\ %
	ISSN 2150-8097. \\
	\href{https://doi.org/\vldbdoi}{doi:\vldbdoi} \\
}\addtocounter{footnote}{-1}\endgroup

\ifdefempty{\vldbavailabilityurl}{}{
	\vspace{.3cm}
	\begingroup\small\noindent\raggedright\textbf{PVLDB Artifact Availability:}\\
	The source code, data, and/or other artifacts have been made available at \url{\vldbavailabilityurl}.
	\endgroup
}
	
\input{introduction}
\input{preliminaries}
\input{SystemOverview}
\input{systemDesign}
\input{evaluation}

\input{related}

\input{conclusion}

%\end{document}  % This is where a 'short' article might terminate

% ensure same length columns on last page (might need two sub-sequent latex runs)
\balance

%ACKNOWLEDGMENTS are optional
%\section{Acknowledgments}
%This section is optional

%\newpage
\bibliographystyle{ACM-Reference-Format}
\bibliography{main}

%APPENDIX is optional.
% ****************** APPENDIX **************************************
% Example of an appendix; typically would start on a new page
%pagebreak

\end{document}

%% file: introduction.tex
\section{Introduction}
\label{sec:introduction}

Business Intelligence (BI) applications allow users to analyze the underlying data using structured queries, but usually rely on a semantic layer, captured as an OLAP cube definition, to organize the data into measures and dimensions\footnote{Measures are quantifiable entities and dimensions are categorical attributes.}. Users rely on predefined dashboards to analyze the data and drive insights. 
To answer questions that are not contained in such predefined dashboards, the missing information needs to be incorporated into the dashboards by technical users, which can be prohibitively expensive and time-consuming, delaying key business insights and decisions. 
Conversational interfaces to BI applications~\cite{DBLP:journals/pvldb/QuamarOMMNK20} have the potential to democratize access to data analysis using a natural language dialog, for non-technical users ranging from top level executives to data scientists. 
Recently, there has been a rapid proliferation of conversational interfaces for data exploration~\cite{AskData,CognosAssistant,PowerBi,quicksight,thoughtspot, bqdataqna} that leverage the advances in the areas of Natural Language Processing (NLP) and Artificial Intelligence (AI). Most of these available systems are often one-directional or user-driven, relying on the user to formulate the natural language query, and their interfaces also often restrict the dialog to fit a fixed set of patterns. Hence, the user consequently needs to have a certain degree of familiarity with the schema to formulate the right queries. While some of these systems help the user complete their current query with a simple list of suggestions, they fall short in guiding the users in their end-to-end data analysis. What is needed is the guidance in terms of recommendations for the next data analysis step that takes into account the current state as captured by the conversational context.

\begin{figure}[t]
    \begin{center}
    \includegraphics[width=0.7\columnwidth]{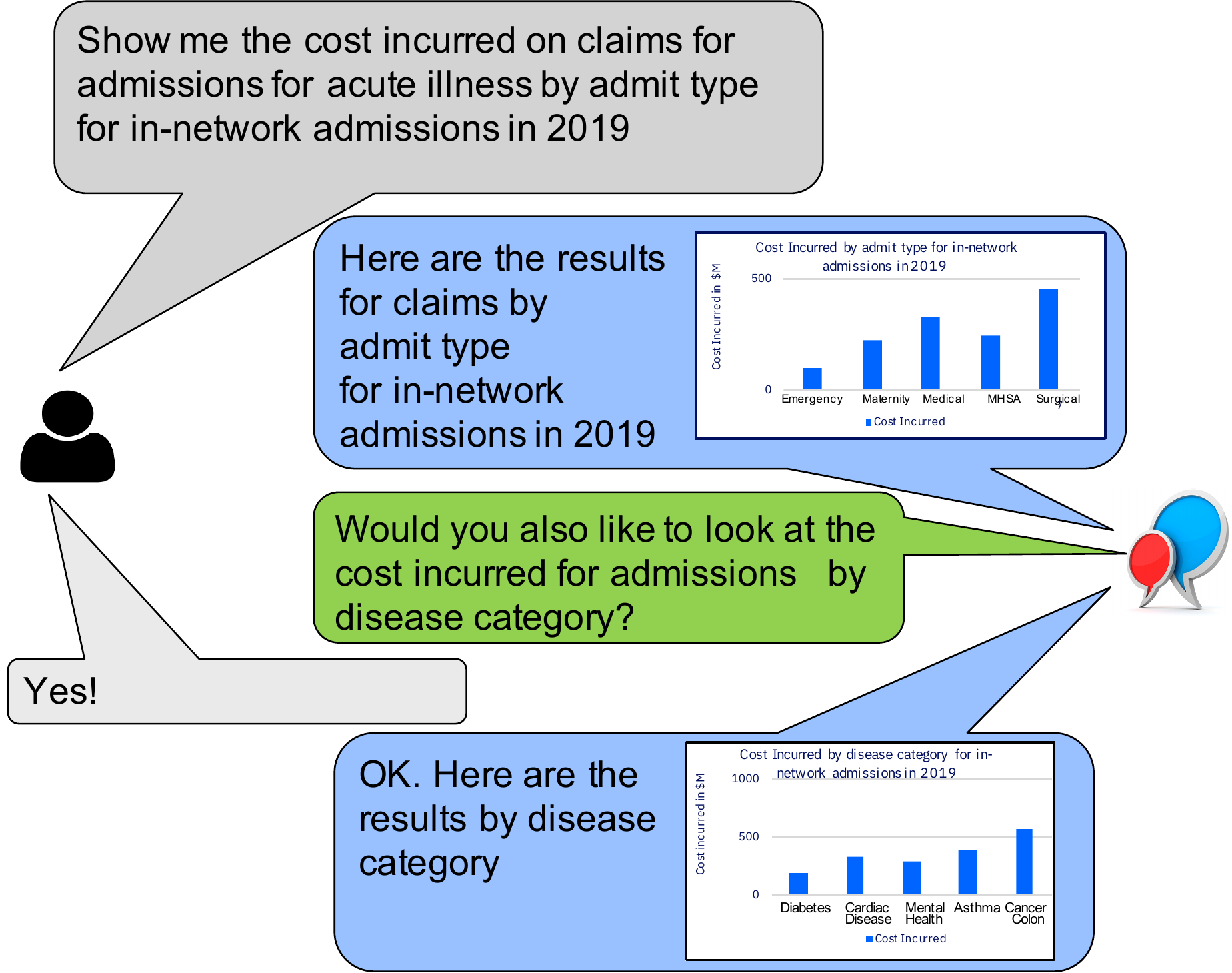} 
    % \vspace{-10pt}
    \caption{Guided data analysis for BI applications.}
    % \vspace{-15pt}
    \label{fig:GuidedConvBI}
    \end{center}
\end{figure}
        
Figure~\ref{fig:GuidedConvBI} shows an example guided conversational data analysis. In this extension, the system is an active participant and provides useful recommendations to guide the user, enabling exploration of data and derivation of insights in small incremental steps. After the first question and response, the system actively recommends to the user to look at the costs incurred on claims by another categorical attribute relevant to admission costs {\em `the disease category'} (shown in green) since the user had originally requested to see the distribution by {\em `admit type'}. In the absence of such recommendations, the user would need to have the knowledge of all possible dimensions that the cost could be sliced/diced by. 
Guided data analysis alleviates the need for the user to understand the data schema and the cube definition, and improves the efficiency and effectiveness of the analysis task by helping the user derive useful insights in fewer iterations. The system actively recommends meaningful next steps, hence the user does not have to ask all the questions.

Enabling guided exploration for BI applications, as in the example of Figure~\ref{fig:GuidedConvBI}, faces several challenges. 
The first challenge is understanding a user's current \emph{state} in the data analysis. This entails keeping track of the queries that the user has issued so far, what subset of data has already been explored, and the user profile, to enable tailored recommendations to different personas. Addressing this challenge requires efficient mechanisms for state management.
Second challenge is the need for deep understanding of the BI application, such as understanding which analysis task a user is interested in, which quantitative and categorical attributes in the underlying data are relevant to the task, and how these are related to each other. 
Third challenge is the navigation of the search space for making recommendations. The search space for a BI analysis query is very large, owing to the combinatorial explosion of the possible combinations of the quantitative and categorical entities in the underlying data and the possible BI operations over them. 
This creates challenges in compact representation of the search space and efficient algorithms to explore the search space for meaningful recommendations.

There is a substantial body of prior work in the area of recommendations for data exploration. \citet{baseline, DBLP:journals/tods/MeduriCS21, zhang-etal-2019-editing} utilize prior workload information against the dataset to recommend SQL queries. They extract patterns from prior workloads and use Machine Learning (ML) techniques to recommend historical queries similar to the current query/session. While~\cite{DBLP:journals/tods/MeduriCS21, zhang-etal-2019-editing} can also recommend unseen SQL queries which may not be present in prior logs, by using SQL synthesis and editing operations respectively, they are still limited by the fact that the user needs to be aware of the underlying database schema to issue the queries, and the schema information is also encoded into the feature vectors assisting the ML model training.
Other related works such as~\citet{DBLP:conf/sigmod/YanH20, DBLP:conf/kdd/MiloS18, DBLP:conf/edbt/SomechMO19} focus on recommending lower level structured query operators/operations, such as select, join, filter and group by, using a combination of ML techniques, interestingness metrics and SQL heuristics.
However, these works do not focus on the higher level BI analysis and have a limited understanding of the common BI patterns or the OLAP cube, as they represent the data analysis operations directly in terms of SQL operators.

In this paper, we propose \sys, a conversational recommendation system for BI applications. At any point in the conversation, \sys, an active participant, recommends a set BI analysis steps, expressed as patterns that are relevant to the user's current state of data analysis. 
We define the search space of possible analysis states in terms of (1) BI patterns, which include a BI operation (such as \texttt{ROLL-UP}, \texttt{DRILL-DOWN}, etc), a measure of interest, as well as dimensions, filters, associated operations and (2) {\em Ontology Neighborhood}, which contains relevant semantic information extracted from the BI ontology. Building on our earlier work \cite{DBLP:journals/pvldb/QuamarOMMNK20}, we exploit a semantic abstraction layer which captures the semantics of the data in terms of measures, dimensions, and their relationships, creating a BI ontology. 
\sys\ utilizes this semantic abstraction layer and prior user interactions with the dataset to provide useful recommendations. 
\sys\ extracts data analysis patterns from prior user sessions and augments the patterns with semantic information from the ontology neighborhoods so that \sys\ not only recommends BI patterns similar to the states that are seen in the query logs, but also unobserved states which are ``close'' to the user's current analysis state.
Further more, \sys\ uses the conversational context to extract information about the current and past queries in the data analysis session as well as any profile information that can assist in tailoring the recommendations to appropriate personas. 

We model a user's data analysis state as a graph that combines information from both BI patterns and ontology neighborhoods. To create a compact representation of the states, we utilize GraphSAGE~\cite{NIPS2017_5dd9db5e}, a Graph Neural Network (GNN) algorithm for inductive representation learning of graphs. With GraphSAGE, we generate low-dimensional vector representations of analysis states that are used by our recommendation algorithms. 

We propose a novel two-step approach to divide and conquer the huge search space of data analysis states. In the first step, we use a multi-class classifier to predict the high level action in terms of a BI operation and the measure of interest, followed by a novel index-based collaborative step that produces full BI pattern recommendations. 
This approach allows us to train multi-class classifier models that have high accuracy with a small amount of training data that focuses on the distinct combinations of BI operation and the measures. Once we identify the high-level action, we can utilize an index that organizes the states based on the high-level action to prune the search space for the subsequent collaborative filtering step. This two-step approach not only results in efficient execution time that is needed for interactive conversational analysis, but also produces highly accurate recommendations, similar to an exhaustive baseline.

We provide a detailed experimental evaluation of our system and have conducted a user study to ascertain the effectiveness of our proposed techniques for BI pattern recommendation. Our experimental evaluation shows that \sys~achieves an accuracy of 83\% for BI pattern recommendations and up to 2$\times$ speedup in latency of prediction over a competitive state-of-the-art baseline. Our user study further indicates that \sys~provides recommendations with a precision@3 of 91.90\% across several different analysis tasks, making it very effective in navigating the search space and providing useful recommendations.

The main contributions of our work are:
\begin{itemize}[leftmargin=*]
    \item An end-to-end system, \sys, for recommending BI queries to a user for guided data analysis through a conversational interface.
    \item Novel techniques to represent a user's current state of data analysis as a graph embedding, capturing both the BI patterns as well as the relevant semantic information extracted from the BI ontology.
    \item A two-step approach to explore the huge search space of possible BI pattern recommendations, which first identifies the high-level action using a multi-class classifier, and then completes the recommendation with a novel indexed collaborative filtering technique.
    \item Detailed experimental analysis and a user study to evaluate the effectiveness of our proposed techniques. 
\end{itemize}
        
The remainder of the paper is organized as follows. Section~\ref{sec:preliminaries} describes our earlier work on conversational BI, including our semantic layer, as well as the terms and definitions used in \sys. Section~\ref{sec:overview} provides an overview of the architecture of \sys, followed by our proposed techniques for state representation and BI pattern prediction in detail in Sections~\ref{sec:stateRepresentation} and~\ref{sec:BIQueryPrediction}. In Section~\ref{sec:evaluation}, we provide  a detailed evaluation of \sys\ including a user study. Section~\ref{sec:relatedWork} explores and compares prior work, and we conclude in Section~\ref{sec:conclusion} with directions for future work.

%% file: preliminaries.tex
\section{preliminaries}
\label{sec:preliminaries}

In this section, first we provide an overview of the key concepts, terminologies and modeling techniques from our earlier work on building a conversational BI system~\cite{DBLP:journals/pvldb/QuamarOMMNK20}.
Next, we introduce how we model prior user interactions in \sys.

\subsection{Health Insights: A Conversational BI System}
\label{sec:HealthInsights}
In our prior work~\cite{DBLP:journals/pvldb/QuamarOMMNK20}, we built a conversational interface to help users analyze a healthcare dataset, called Health Insights (HI), using natural language dialog. The Health Insights{\sffamily\textregistered}~ product is an IBM{\sffamily\textregistered}~ Watson Health{\sffamily\textregistered}~ offering~\cite{HealthInsights} that includes several curated datasets of healthcare insurance data related to claims and transactions from a population covered by an insurance's healthcare plans. We exploit the OLAP cube definition against the HI dataset to learn a semantically rich entity-centric view of the underlying BI schema called the {\it Semantic Abstraction Layer (SAL)}. We use the semantic information in SAL to bootstrap the conversation system with the relevant entities and relationships. We also identify the common access patterns for BI analysis, and use them to interpret the user's utterance and generate structured SQL queries against the database. In the following subsections, we will shortly describe the dataset, the semantic abstraction layer, and the BI patterns, as we make use of the SAL and the BI patterns heavily in ~\sys. More details can be found in ~\cite{DBLP:journals/pvldb/QuamarOMMNK20}.

\subsubsection{HI Dataset}
\label{sec:preliminaries:data}
The HI dataset consists of basic information about participants' drug prescriptions and admissions, service, key performance factors such as service categories, as well as individual patient episodes\footnote{Patient episodes are a collection of claims that are part of the same incident to treat a patient.}.
In addition, the HI dataset also includes the IBM MarketScan{\sffamily\textregistered}~ dataset~\cite{MarketScan} contributed by large employers, managed care organizations and hospitals. The dataset contains anonymized patient data including medical, drug and dental history, productivity including workplace absence, laboratory results, health risk assessments, hospital discharges and electronic medical records (EMRs).

\subsubsection{Semantic Abstraction Layer (SAL)} 
\label{sec:preliminaries:semantic}
We capture the OLAP cube definition over the HI data in the form of an ontology, which we call the {\em BI Ontology}. A BI ontology provides a semantically rich and entity-centric view of the BI schema in terms of quantifiable entities called {\em Measures}, categorical attributes called {\em Dimensions}, their hierarchies and relationships as defined in the OLAP cube definition. Each measure and dimension described in the OLAP cube definition is represented as a class in the BI ontology and annotated as an actual measure/dimension. The measure and dimension hierarchies captured from the OLAP cube definition are represented as functional relationships in the BI ontology. For example, in a dimensional hierarchy for time, each of the time dimensions such as year, month, week, and day would be connected using directed edges representing functional relationships between the time dimensions.

\begin{figure}[t]
\begin{center}
\includegraphics[width=\columnwidth]{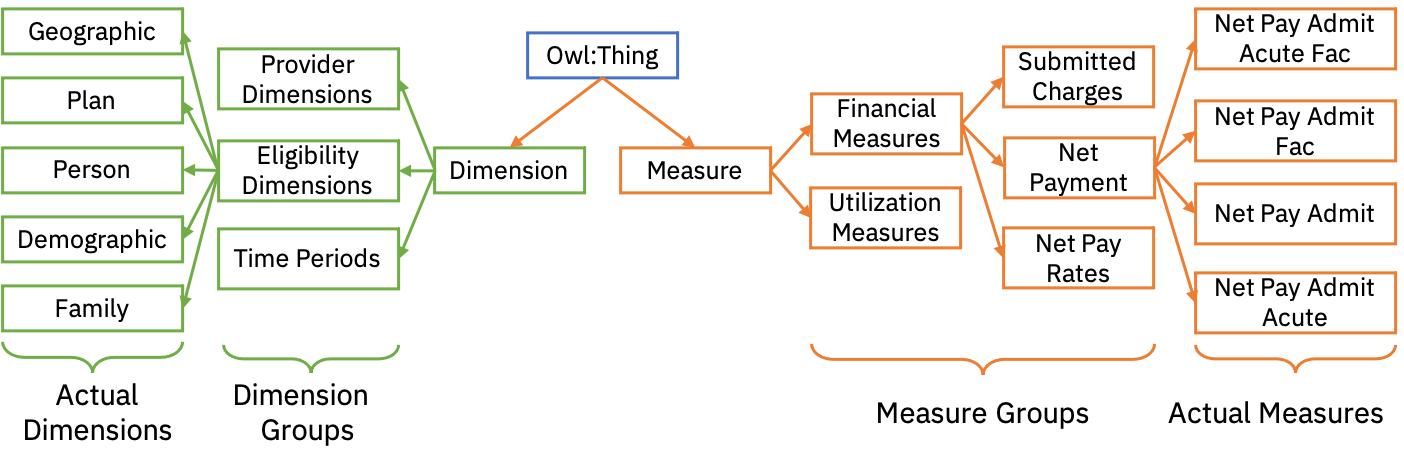} 
%\vspace{-10pt}
\caption{BI ontology: measure and dimension grouping.}
\vspace{-15pt}
\label{fig:measureGrouping}
\end{center}
\end{figure}

We further augment the BI ontology with higher-level logical grouping of measures, called {\em Measure Groups (MGs)}, and of dimensions, called {\em Dimension Groups (DGs)}. This grouping is provided by subject matter experts (SMEs), to enable the HI system to better understand the analysis task and the dataset.
This way, our system can efficiently navigate to relevant portions of the underlying BI schema to determine the measures and dimensions that are relevant to the analysis task. Figure~\ref{fig:measureGrouping} shows one possible grouping of measures and dimensions in the augmented BI ontology. In Figure~\ref{fig:measureGrouping}, {\em Net Pay Admit} is an actual measure defined in the OLAP cube over the underlying data, and that {\em Net Payment} is a logical grouping provided by SMEs. 

Each logical grouping of measures and dimensions provided by the SMEs is also represented as a class and is annotated as a measure/dimension group respectively in the BI ontology. Measure and dimensions are grouped into groups using {\em is-A} (parent-child) relationships in the BI ontology. Note that some real-world applications and datasets may not have these higher-level logical groupings of measures and dimensions in terms of measure/dimension groups. \sys~uses the measure and dimension groups, if they are available, otherwise it uses the original BI ontology derived from the OLAP cube definition.

\subsubsection{Modeling BI Patterns}
\label{sec:BI_Patterns}
Each user utterance is characterized by well-structured {\em BI Patterns}. The constituent elements of each BI pattern are discerned from the natural language queries using a trained classifier and NLP techniques such as Named Entity Recognition (NER) employed by the conversational interface. 

We define a BI Pattern ($P_{BI}$)~\cite{DBLP:journals/pvldb/QuamarOMMNK20} as a quadruple (Equation~\ref{eqn:BI_Pattern}) consisting of (1) $op_{BI}$, a BI-specific operation from a set of operations $OP_{BI}$ = $\{$\texttt{ANALYSIS}, \texttt{DRILL-DOWN}, \texttt{ROLL-UP}, \texttt{PIVOT}, \texttt{TREND}, \texttt{RANKING}, \texttt{COMPARISON}$\}$, (2) $M$, a set of measures (or measure groups) defined in the BI ontology, (3) $D$, a set of dimensions (or dimension groups) defined in the BI ontology and (4) $O_{Query}$, a set of query operations such as \texttt{AGGREGATION} on measures, \texttt{GROUP BY} and \texttt{FILTER} on dimensions.
\begin{equation}
    P_{BI} = <op_{BI}, M, D, O_{Query}>\\
\label{eqn:BI_Pattern}
\end{equation}
where $op_{BI} \in OP_{BI}$. We provide an example BI pattern below and refer the reader to~\citet{DBLP:journals/pvldb/QuamarOMMNK20} for further details.

\begin{comment}
\begin{figure}[!ht]
    \begin{center}
    \includegraphics[width=\columnwidth]{figures/BIAnalysisPattern.pdf} 
    %\vspace{-15pt}
    \caption{BI analysis pattern.}
    %\vspace{-10pt}
    \label{fig:analysisPattern}
    \end{center}
\end{figure}
\end{comment}
\begin{comment}
\begin{example}
The {\em BI analysis query pattern} is one of the most common BI patterns observed in BI workloads. It allows users to analyse a measure sliced along a particular dimension, optionally with a filter. 
Figure~\ref{fig:analysisPattern} shows an example of this pattern, where the user is enquiring about the number of {\em admits} (measure) by {\em Major Diagnostic Category} (dimension) for the year {\em 2017} (a filter value). The pattern can be represented as the quadruple: $P_{BI}$ = $<$\texttt{ANALYSIS}, \{{\em Admits}\}, \{{\em Major Diagnostic Category}\}, \{\texttt{COUNT}, \texttt{YEAR}={\em 2017}\}$>$
\end{example}
\end{comment}

\begin{figure}[!ht]
    \begin{center}
    \includegraphics[width=\columnwidth]{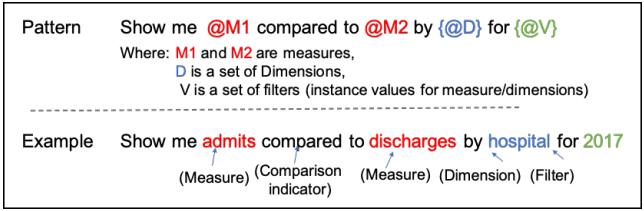} 
    %\vspace{-15pt}
    \caption{BI comparison pattern.}
    %\vspace{-10pt}
    \label{fig:comparisonPattern}
    \end{center}
\end{figure}

\begin{example}
A common BI pattern observed is the {\em BI comparison pattern} which allows users to compare two or more measures against each other along a particular dimension and optionally with a filter value. Figure~\ref{fig:comparisonPattern} shows an example BI comparison pattern that compares the number of {\em admits} to {\em discharges by hospital} (dimension) for the year {\em 2017} (a filter value). The pattern can be represented as the quadruple: $P_{BI}$ = $<$\texttt{COMPARISON}, \{{\em Admits, Discharges}\}, \{{\em Hospital}\}, \{\texttt{COUNT}, \texttt{YEAR}={\em 2017}\}$>$
\end{example}

\subsection{Modeling Prior User Interactions for \sys}
\label{sec:preliminaries:user}
Conversational logs that capture prior user interactions against a data set are a rich source of information. Analysis of these logs enables extraction of useful data analysis patterns. \sys~effectively leverages these patterns to inform query recommendations for current user interactions against the data set. Here, we describe how we model these prior user interactions and provide definitions for the key terms and concepts used in the paper.

Prior user interactions are characterized by a sequence of NL queries issued by the user and corresponding responses provided by the system across several conversational turns\footnote{A conversational turn is a pair consisting of a user utterance (or query) and the system response to the user utterance.}. We capture the conversational logs of prior user interactions in terms of the following:

%\textit{Query.} 
A \textit{query} is the natural language question/utterance issued by the user at a given state of data analysis\footnote{We use the term {\em data analysis state} and {\em state} interchangeably in the paper.}. Each query is interpreted as a BI pattern $P_{BI}$, along with its constituent elements defined in Section~\ref{sec:preliminaries:semantic}. Further, each $P_{BI}$ is translated into a SQL query called {\em BI Query}, issued against the database to retrieve the results for the user query. 

%\textit{State.} 
A \textit{state} ($S$) represents the context of data analysis in terms of (1) the BI pattern $P_{BI}$, including its constituent elements extracted from the query issued by the user, (2) the measure group the user is interested in, and (3) the elements from the BI Ontology that are relevant to $P_{BI}$, allowing for flexibility in making recommendations in terms of unseen but similar queries. Section~\ref{sec:stateRepresentation} describes this in more detail.

%\textit{User Session.} 
A \textit{user session} ($US$) is a sequence of states capturing the analysis done by the user in a single sitting.
We model $US$ as a simple linear graph, wherein each node in the graph represents a state. Each directed edge between two states (a source state and a target state) represents a query issued by the user at the source state  to reach the target state. The first and the last states in each user session are termed {\em Initial State} and {\em Final State}, respectively.

\begin{figure}[!ht]
    \begin{center}
    \includegraphics[width=\columnwidth]{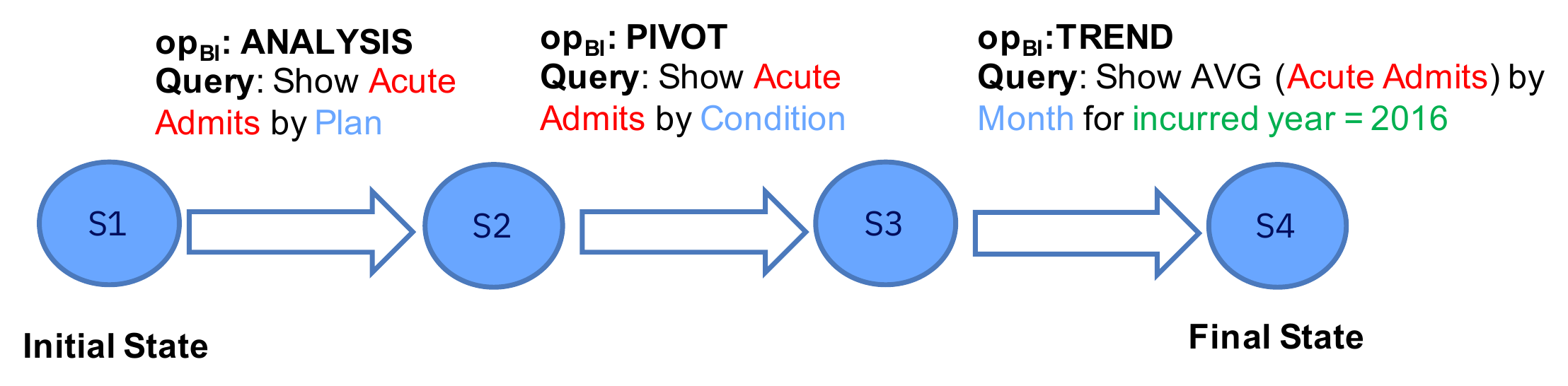} 
    %\vspace{-10pt}
    \caption{A user session example.}
    %\vspace{-15pt}
    \label{fig:userSession}
    \end{center}
\end{figure}

Figure~\ref{fig:userSession} shows an example data analysis user session obtained from the HI conversational logs. The session is represented as a sequence of four states and three queries representing a user's transition from an initial state of data analysis $S_1$, to a final state of data analysis $S_4$. 
For each natural language query issued by the user at a particular state, the system identifies $P_{BI}$ associated with the query and extracts all the relevant features required for populating the state including the $op_{BI}$ (e.g., \texttt{ANALYSIS, PIVOT, TREND}), measure (e.g., {\em Acute Admits}), dimensions (e.g., {\em Plan}, {\em Condition}, {\em Month}) and filters (e.g., {\em Incurred Year = 2016)} as shown in Figure~\ref{fig:userSession}. 

%\textit{Session Task.} 
In addition to the feature extraction for each state, we also annotate each user session $US_k$, with a \textit{session task} $Task_{US_{k}}$. $Task_{US_{k}}$ represents the semantically higher-level information that the user is interested in analyzing in the session. BI analysis is typically characterized by users looking at a specific measure(s) which they slice and dice along several dimensions and their hierarchies using different operations $op_{BI}$ to gain useful insights. For example, {\em Acute Admits} is the queried measure for states $S_2$, $S_3$ and $S_4$, as shown in Figure~\ref{fig:userSession} and is the most representative of the analysis task that the user is interested in. We therefore define the session task in terms of the measures queried in the different states of the session.

More specifically, we define $Task_{US_{k}}$ as the union of the parent of each measure (is-A relationship) being investigated in the session (Equation~\ref{eqn:session_task}). We chose the session task to be the immediate parent of the measures being investigated in the session. This affords an appropriate balance between (a) Generalization: providing an intuition of the semantically higher-level information that the user is looking for, and (b) Specialization: being specific enough to the measure(s) that the user is interested in analyzing.
\begin{equation}
     Task_{US_k} = \bigcup_{i=1}^{n} Parent(m_{S_i})
    \label{eqn:session_task}
\end{equation}

For example, {\em Utilization} is the parent of {\em Acute Admits} and defined as a {\em Measure Group (MG)}, a logical grouping provided by the SMEs in the BI ontology.
Hence, it is the session task signifying that the user is interested in analyzing the utilization of health care resources in terms of the admits for acute conditions in the current session. (Section~\ref{sec:stateRepresentation} for further details). 
The function $Parent(m_{S_i})$ returns the immediate MG associated with $m_{S_i}$ if it exists in the BI ontology. If not, it returns the measure $m_{S_i}$ itself\footnote{This could be either because there exists no $Parent(m_{S_i})$ in the cube definition or it is not provided by the SMEs.}. The session level task in this case would thus degenerate to the union of all measures explored by the user in the session.

\subsection{Problem Definition}
\label{sec:preliminaries:problem}
%Having described the necessary notions used, we now proceed with the formal definition of the problem studied in this work.
We define the problem of conversational BI recommendations as follows:
\begin{definition}
Given a conversational log of prior user sessions against a dataset and a BI ontology derived from the cube definition against this dataset, provide top-$k$ BI pattern ($P_{BI}$) recommendations at each state to help the user achieve his/her current analysis goal.
\end{definition}

%% file: SystemOverview.tex
\section{System Overview}
\label{sec:overview}

\begin{figure}[!ht]
    \begin{center}
    \includegraphics[width=\columnwidth]{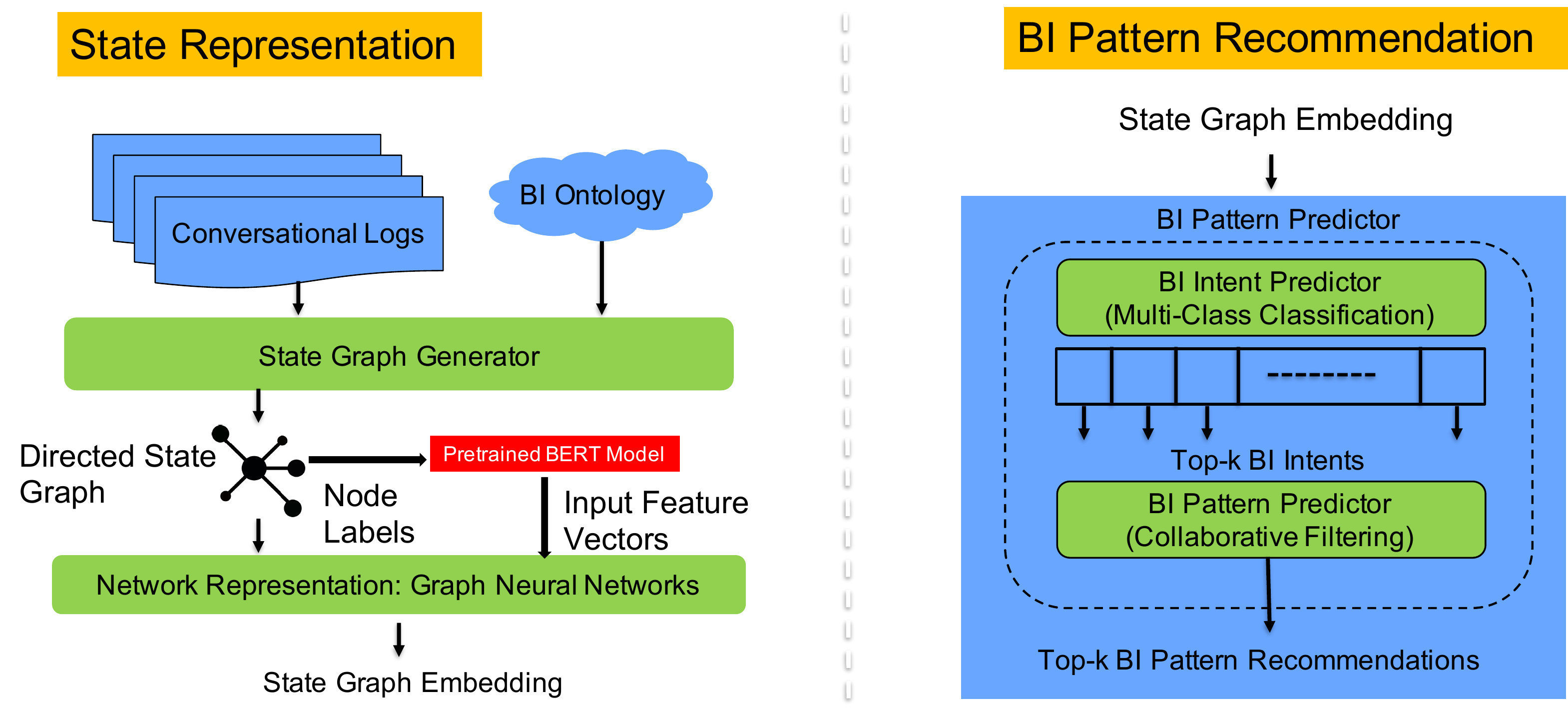} 
    %\vspace{-10pt}
    \caption{\sys\ architecture.}
    %\vspace{-15pt}
    \label{fig:systemArchitecture}
    \end{center}
\end{figure}

This section provides an overview of the architecture of \sys. It consists of (1) an offline {\em State Representation Learning} phase that trains a model to learn a low-dimensional vector representation of each state in a user session and (2) an online {\em BI Pattern Recommendation} phase which takes the latent vector representation of a state (created by the trained model) from a current user session as input and provides the top-$k$ BI pattern recommendations at that particular state of data analysis. Figure~\ref{fig:systemArchitecture} shows the two phases of \sys's architecture.

In the offline phase, the model for creating the state representation is trained using prior user sessions in the conversational logs enriched by the semantic information captured in the BI ontology.

First, for each state in a user session, the state graph generator creates a directed graph that captures the state information learnt from the conversational logs in terms of the BI pattern and its constituent elements. The graph is then further enriched with the session-level analysis task $Task_{US_k}$ and additional semantic information relevant to the entities in the state graph from the BI ontology. The enriched representation of the state allows \sys~to recommend BI patterns similar to the states that are seen in the query logs, but also unseen states which are semantically “close” to the user’s current analysis state (Section~\ref{sec:stateRepresentation}).

The next step is network representation learning for generating the state the graph embeddings (Section~\ref{sec:representationLearning}). A pre-trained language model (BERT)~\cite{devlin-etal-2019-bert} is used to generate fixed-length feature vectors for each node in the state graph using their node labels. The feature vectors (initial node embeddings) are provided as input along with the directed state graph to train a model using GraphSAGE~\cite{hamilton2017representation}, an inductive representation learning framework for graphs. The model captures each state $S_i$ in the form of a low-dimensional vector (i.e., state graph embedding) $E_{S_i}$. This embedding provides a compact representation of features from both the conversational logs as well as the semantic knowledge from the BI Ontology and preserves the structural relationships between the different entities in the state graph.

The online phase of BI Pattern prediction  (Figure~\ref{fig:systemArchitecture}) generates the top-$k$ BI pattern recommendations at each step of an active user session. The search space of BI pattern recommendation (Equation~\ref{eqn:search_space}) is huge, being the Cartesian product of the possible BI operations $OP_{BI}$, measures $M$, dimensions $D$, as defined in the OLAP cube definition, and operations $O_{Query}$ on measures (\texttt{AGGREGATION}) and dimensions (\texttt{GROUP BY} or \texttt{Filter}). 
\begin{equation}
    \mathcal{S} = OP_{BI} \times M \times D \times O_{Query}
    \label{eqn:search_space}
\end{equation}
To divide and conquer this huge search space, \sys\ takes a two-step approach that obviates the need for prediction of the entire BI pattern in one shot. The first step takes the graph embedding of the current state in an active data analysis session as input and predicts a coarse-grained high-level action called BI Intent ($Intent_{BI}$) using a trained multi-class classifier model. Each predicted BI Intent $Intent_{BI}$, is defined as a tuple (Equation~\ref{eqn:BI-Intent}) consisting of the next BI operation $op_{BI} \in OP_{BI}$ (e.g., \texttt{DRILL-DOWN}, \texttt{ROLL-UP}, \texttt{PIVOT}, etc.), and a $MG \in Task_{US_k}$ (e.g., {\em Utilization}, {\em Net Payment}) that the user is interested in the current session. 

\begin{equation}
    Intent_{BI} = \left<op_{BI}, MG\right>
    \label{eqn:BI-Intent}
\end{equation}

Predicting the next data analysis step in terms of an $Intent_{BI}$ helps to significantly narrow down the search space. As seen from Equation~\ref{eqn:search_space_BI_Intent}, the search space for $Intent_{BI}$ prediction $\mathcal{S}_{Intent_{BI}}$, is the Cartesian product of the number of BI operations $OP_{BI}$ and the distinct number of session tasks $Task_{session}$, which is orders of magnitude smaller than the search space $\mathcal{S}$ for BI pattern prediction. This allows \sys\ to train a highly accurate prediction model with a small amount of labelled training data (Section~\ref{sec:BIIntentPrediction}) that is usually expensive to obtain. 
\begin{equation}
    \mathcal{S}_{Intent_{BI}} = OP_{BI} \times Task_{US}
    \label{eqn:search_space_BI_Intent}
\end{equation}

The second step refines the $Intent_{BI}$ into a more detailed BI pattern $P_{BI}$, with all its constituent elements using a novel index-based collaborative filtering  approach ($CF_{Index}$). Using the novel $CF_{Index}$ approach gives \sys~the distinct advantage of producing predictions with an accuracy almost equivalent to an exhaustive collaborative filtering approach while providing significant improvement in terms of lowering prediction latency, a critical requirement for real-time interactions in conversational BI systems. Finally, these top-$k$ BI pattern predictions are further refined by a post-processing step to enhance the quality and richness of the recommendations. 

%% file: systemDesign.tex
\section{State Representation}
\label{sec:stateRepresentation}
Selection of features for representing a state in a user session has a direct impact on the search space of making recommendations. Limiting the features to the information contained in each state extracted from the user sessions in the conversation logs such as BI patterns $P_{BI}$ would restrict the recommendations to similar states seen in the prior user sessions. We propose a novel technique for enrichment of the features for state representation with semantically rich information from the BI Ontology relevant to the current state. This provides a powerful mechanism to expand the search space of our recommendations to states that are semantically similar to the current state but that might not have been seen in prior user sessions. A graph structured representation of each state allows us to meaningfully combine features from both the conversational logs as well as the relevant semantic information from the BI Ontology while preserving the structural relationships between the different entities combined.

\subsection{Graph Structured State Representation}
\label{sec:stateGraphRepresentation}
While creating the structured representation of a state, we start with the information contained in each state as extracted from the user sessions in the conversation logs. This information is limited to the BI pattern $P_{BI}$ observed for the state $S_i$ in a prior user session $US_k$, including the BI operation $op_{{BI}_{S_i}}$, measures $m_{S_i}\in M$, dimensions $d_{S_i}\in D$, and query operations $O_{S_i}$ on these entities (e.g., \texttt{AGGREGATION}, \texttt{FILTER}, etc.).

We further enrich the extracted state information with features extracted from the BI ontology that are semantically relevant to $m_{S_i}$, $d_{S_i}$. These features are termed as the {\em Ontology Neighborhood} $ON_{S_i}$ (Equation~\ref{eqn:ontology_neighborhood}) relevant to the state $S_i$. More specifically, $ON_{S_i}$ consists of the session task $Task_{US_k}$, where $S_i \in US_k$, {\em Expanded Measures} $EM_{S_i}$, which are sibling measures, i.e., children of the measure groups in session task $Task_{US_k}$, and {\em Expanded Dimensions} $ED_{S_i}$, which are dimensions connected to $m \in EM_{S_i}$ in the BI ontology via an edge $e \in E$. 
\begin{equation}
    ON_{S_i} = Task_{US_k} \cup EM_{S_i} \cup ED_{S_i},
   \label{eqn:ontology_neighborhood}
\end{equation}
where
\begin{equation}
    EM_{S_i} = \{m \in M | m = Sibling(m_{S_i})\},
    \label{eqn:expanded_measures}
\end{equation}
\begin{equation}
    ED_{S_i} = \{d \in D |m \in EM_{S_i}, (m,d) \in E \}.
    \label{eqn:expanded_dimensions}
\end{equation}

\begin{figure}[!ht]
    \begin{center}
    \includegraphics[width=\columnwidth]{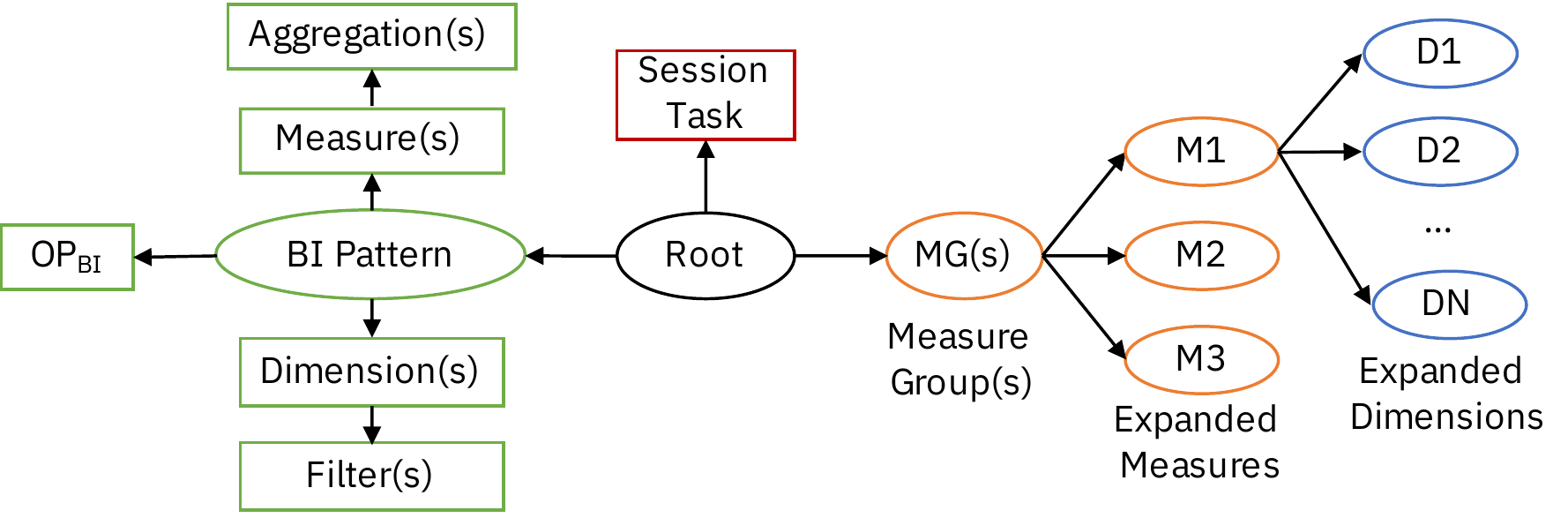} 
    %\vspace{-10pt}
    \caption{Graph structured state representation.}
    %\vspace{-15pt}
    \label{fig:graphRepresentation}
    \end{center}
\end{figure}

We employ a state graph generator module (Figure~\ref{fig:systemArchitecture}) to create the state graph representation for each state. In a state graph (Figure~\ref{fig:graphRepresentation}), each node represents the extracted state features, and an edge represents the relationships between them. The edges are directed to represent the structural dependency of the features within the state. For instance, operations such as \texttt{AGGREGATIONs} often co-occur with measures, and \texttt{FILTERs} co-occur with dimensions in a query. This necessitates an edge between each measure and its associated aggregation. Similarly, there is an edge between a filter operation and the dimension to which it is applied. Each of the nodes representing the $op_{BI}$, measures $m_{S_i}$, dimensions $d_{S_i}$ are also connected with an edge to the BI Pattern node that together provide a structured representation the query issued by the user. For the nodes representing the ontology neighborhood $ON_{S_i}$, the directed edges between measures and measure groups, dimensions and dimension groups denote hierarchical (is-A) relationships. Edges between expanded measures $EM_{S_i}$ and expanded dimensions $ED_{S_i}$ represent functional relationships. 
The graph also has a root node that is artificially introduced and connected via separate edges to the BI pattern and measure group nodes. The root node is associated with one attribute, the session task $Task_{US_k}$ extracted from the BI ontology.

\subsection{Representation Learning on State Graphs}
\label{sec:representationLearning}

Having constructed the state graph, we now describe our representation learning technique to generate state graph embeddings. We use GraphSAGE~\cite{NIPS2017_5dd9db5e}, an inductive representation learning framework, to create a vectorized representation of the state graphs in the form of graph embeddings. The key reasons of utilizing GraphSAGE are described below.

\textbf{Inductive unsupervised setting.} This setting extends Graph Convolution Networks (GCNs) to learn aggregator functions as embedding functions that can generalize to unseen nodes~\cite{NIPS2017_5dd9db5e}. This key feature enables generalization across state graphs and hence graph embeddings can be generated for unseen state graphs. This inductive setting is critical as it allows us to compute the graph embeddings of new states seen in active user sessions which could then be used as input to downstream prediction models for recommendations.

\textbf{Learning from the neighborhood.} 
This allows the embedding to learn from both the structure (global information) and node features (local information) of the elements in the state graph, both of which are necessary to capture the semantic information represented by a state in a user session. Finally, the ability to learn across several layers (multiple hops) allows aggregation of information across the entire state graph.

\begin{figure}[h]
    \begin{center}
    % \vspace{-15pt}
    \includegraphics[width=\columnwidth]{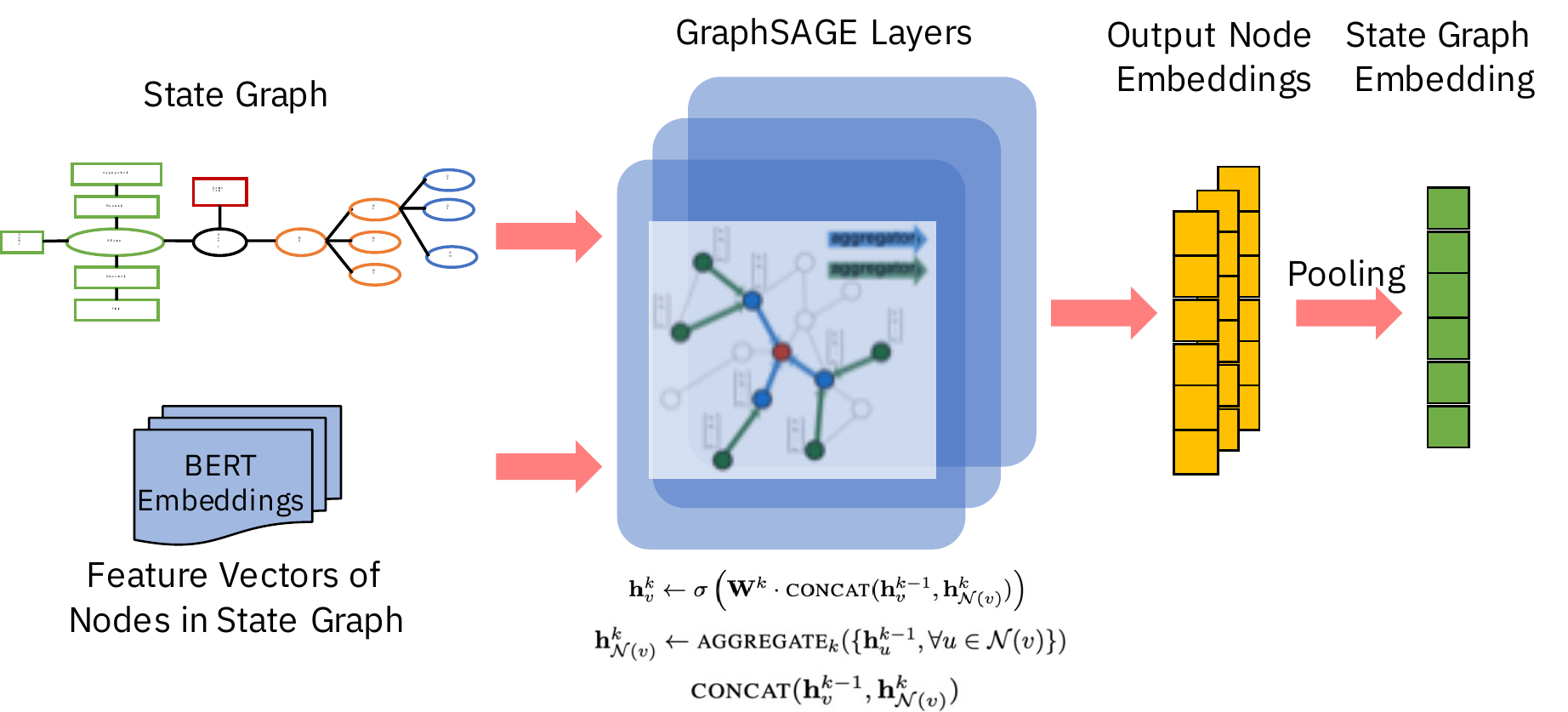} 
    %\vspace{-10pt}
    \caption{State graph representation learning network.}
    %\vspace{-15pt}
    \label{fig:representationLearning}
    \end{center}
\end{figure}

Figure~\ref{fig:representationLearning} shows the end-to-end state graph representation learning network using GraphSAGE for generating state graph embeddings. We first describe the embedding generation process using a forward propagation algorithm which assumes that the model has already been trained. Next, we describe how we train the model.
 
\subsubsection{State Graph Embedding Generation} 
The network takes as input the state graph and a set of input feature vectors for each node in the graph. We use a pre-trained language model BERT~\cite{devlin-etal-2019-bert} to generate node embeddings as input feature vectors corresponding to the names of the nodes (node labels) in the state graph~\footnote{Node labels represent the names of the measures, dimensions, their hierarchies, $op_{BI}$, query operations $O_{S_i}$ that the nodes in the state graph represent.} (Figure~\ref{fig:graphRepresentation}). The embedding generation process involves aggregation of local neighborhood information over several GraphSAGE layers. In each iteration every node $\nu$ first aggregates the node features from its immediate neighbors recursively into a single aggregated feature vector $h^k_{N(v)}$ (Equation~\ref{eqn:aggregation}) and then concatenates its current feature vector $h^{k-1}_v$ with the aggregated feature vector $h^k_{N(v)}$. There are several choices of aggregation operations (MEAN/LSTM/MaxPool) for aggregating neighborhood features. In our current implementation we use MEAN as the aggregator function, following the empirical observation from~\citet{NIPS2017_5dd9db5e}. In addition, the MEAN pooling function gives equal weightage to all the neighborhood features by computing the mean over all of them. This concatenated feature vector is then fed through a fully connected network with a non-linearity $\sigma$ to finally produce $h^k_v$ (Equation~\ref{eqn:concatenation}) that is then fed to the next layer as input. 

\begin{equation}
\label{eqn:aggregation}
\mathbf{h}^{k}_{\mathcal{N}(v)} = \mathrm{AGGREGATE}({\mathbf{h}^{k-1}_{u},\forall u\in \mathcal{N}(v)})
\end{equation}

\begin{equation}
\label{eqn:concatenation}
\mathbf{h}^{k}_{v} = \sigma(\mathbf{W}^k \cdot \mathrm{CONCAT}(\mathbf{h}^{k-1}_v,\mathbf{h}^k_{\mathcal{N}(v)})
\end{equation}

After $k$-layers\footnote{We have set $k$ to 3 for computing the state graph embeddings to ensure that node features from all nodes are propagated and aggregated into the root node.}, we pool the node embeddings of all the nodes in the state graph to create the state graph embeddings.

\subsubsection{Representation Network Model Training} 
We now describe how we train the representation learning network in an unsupervised setting. To generate training data, we randomly sample pairs of states from the training set of user sessions and use an unsupervised loss function (Equation~\ref{eqn:lossFn}) that minimizes the difference between the graph similarity (i.e., {\em Jaccard} similarity) of the pairs in the original space and the latent similarity (i.e., cosine similarity) in the vector space. The {\em Jaccard} similarity treats individual components of the graph as sets to compute their similarity. In Equation~\ref{eqn:JaccardSimilarity}, we include the BI operation $op_{BI}$, measures $m_{s_i}$ with \texttt{AGGREGATIONs}, \texttt{GROUP-BY}, and dimensions $d_{S_i}$ with \texttt{FILTERs} from the BI queries as well as the ontology neighborhoods, $ON_{S_i}$ (i.e., expanded measures and dimensions) in the pair of graphs.
\begin{equation}
\begin{split}
    Sim(S_i, S_j) = \mathrm{AVG}(&JaccSim(op_{{BI}_{S_i}}, op_{{BI}_{S_j}} ),\\
    &JaccSim(m_{s_i}, m_{s_j}), JaccSim(d_{S_i}, d_{S_j}),\\
    &JaccSim(ON_{S_i}, ON_{S_j}))
    \end{split}
    \label{eqn:JaccardSimilarity}
\end{equation}

Our Jaccard similarity gives more weightage to the BI elements of a BI pattern as compared to the measure groups and expanded measures derived from the ontology neighborhood $ON_{S_i}$. The reason is that the elements of a BI pattern are actually extracted from user queries in NL and hence should get a higher weightage than the ontology neighborhood that is proximal to the queried BI elements in the ontology graph. As shown in Equation~\ref{eqn:JaccardSimilarity}, the ontology neighborhood counts only 25$\%$ weightage (1 out of 4 terms), whereas 3 out of 4 terms correspond to the elements in a BI pattern. While alternative set similarity or graph edit-distance based similarity metrics can be used to compute the graph similarity, we have empirically observed a competent accuracy of around 90\% for state representation in Section~\ref{sec:stateEval} using Jaccard similarity.

The loss is then back propagated from the output layer to train GraphSAGE. Equation~\ref{eqn:lossFn} provides the mathematical representation of the loss minimization objective function:
\begin{equation}
min \sum_{i,j \in Pairs} |Sim(S_i,S_j)-CosineSim(V_i,V_j)|
\label{eqn:lossFn}
\end{equation}
where $i$ and $j$ denote indices of the states $S_i$ and $S_j$ in a randomly drawn matching (positive sample) or non-matching (negative sample) state pair from the Cartesian product of state pairs, denoted by “Pairs”. $V_i$ and $V_j$ denote the latent vectors (i.e., graph embeddings) of $S_i$ and $S_j$, respectively. The objective function minimizes the cumulative difference between the Jaccard similarity and the cosine similarity over all such pairs selected in the training set.

\section{BI Pattern Prediction}
\label{sec:BIQueryPrediction}
In this section, we describe in detail the online phase of \sys\ that generates the top-$k$  recommendations in terms of BI patterns $P_{BI}$ at each step of data analysis. We employ a novel two-step approach to divide and conquer the huge search space $\mathcal{S}$ (Equation~\ref{eqn:search_space}), of predicting the next BI pattern in a current user session. The two-step approach first predicts a $Intent_{BI}$ (Equation~\ref{eqn:BI-Intent}) which has a much smaller search space for prediction  (Equation~\ref{eqn:search_space_BI_Intent}). This enables \sys~to train and employ a highly accurate model with a small amount of training data for $Intent_{BI}$ prediction. Subsequently, the $Intent_{BI}$ is expanded into a $P_{BI}$ with all its constituent elements using an efficient index-based collaborative filtering approach $CF_{Index}$ to provide the top-k BI pattern $P_{BI}$ recommendations in real-time.

\begin{figure}[!ht]
    \begin{center}
    \includegraphics[width=\columnwidth]{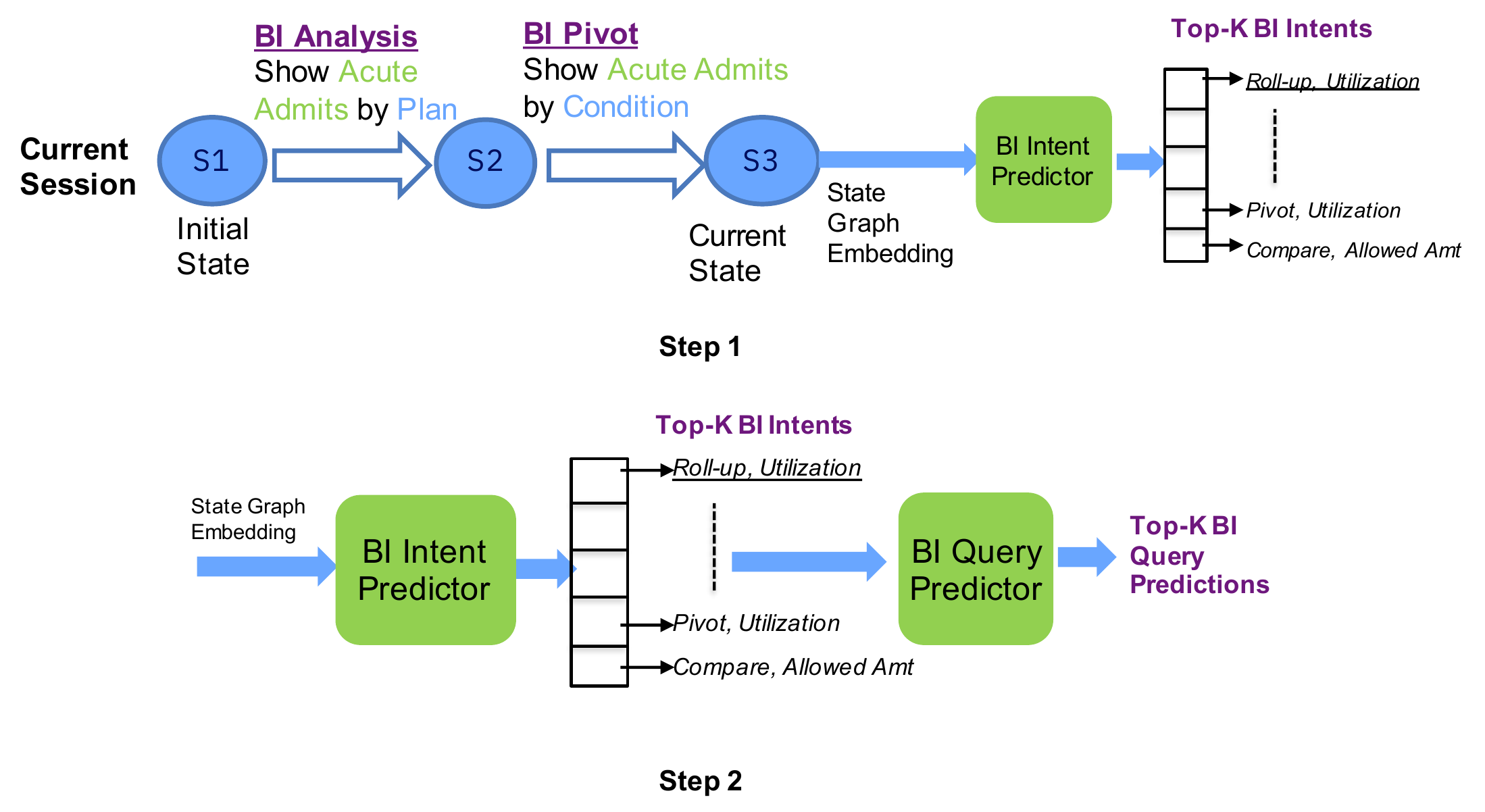} 
    %\vspace{-15pt}
    \caption{Two-step approach for BI query prediction.}
    %\vspace{-15pt}
    \label{fig:twoStepApproach}
    \end{center}
\end{figure}

\subsection{Step1: Top-k BI Intent Prediction}
\label{sec:BIIntentPrediction}
\begin{figure}[!ht]
    \begin{center}
    \includegraphics[width=0.75\columnwidth]{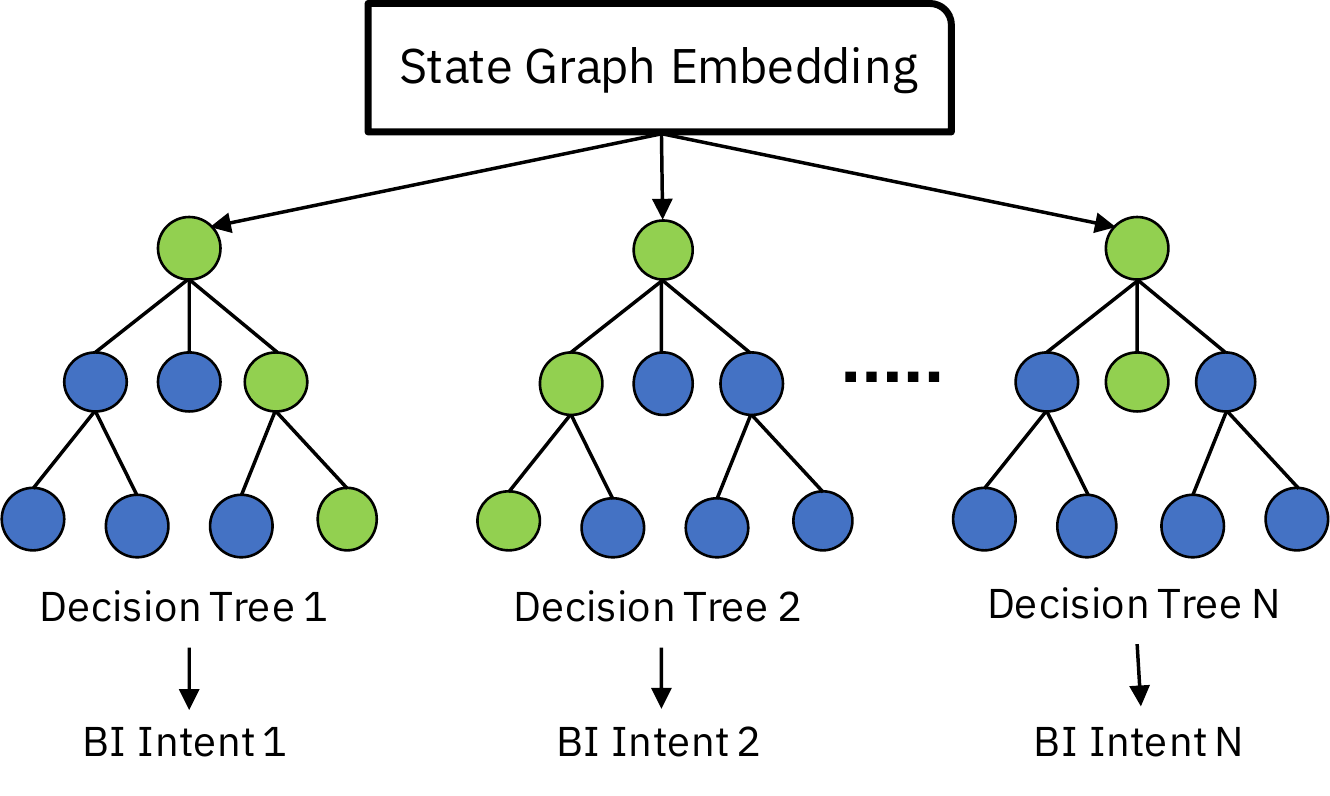} 
    %\vspace{-15pt}
    \caption{Random Forest classifier for BI intent prediction.}
    %\vspace{-10pt}
    \label{fig:RF}
    \end{center}
\end{figure}

We model the $Intent_{BI}$ prediction as a multi-class classification problem that takes the current state graph embedding $E_{S_i}$ as input and provides the top-$k$ $Intent_{BI}s$ as output (Ref Figure~\ref{fig:twoStepApproach}). We trained and employed a Random Forest (RF) classifier (Figure~\ref{fig:RF}) as a $Intent_{BI}$ predictor.
The RF classifier takes $E_{S_i}$ as input and enables ensemble learning across a set of decision trees. 
The RF classifier is trained using labeled examples of $\left<E_{S_i}, Intent_{BI}\right>$ pairs drawn from the conversational logs and a sparse categorical cross-entropy loss function. Each $Intent_{BI}$ represents a distinct class for which the RF classifier emits a probability score for each input test embedding during the prediction phase. \sys\ then chooses the top-$k$ most likely $Intent_{BI}$s based on the probability scores.

In addition to the RF classifier we also implemented alternative models for the multi-class classifier for predicting the top-$k$ $Intent_{BI}$s including: (1) a LSTM classifier which captures the sequence of states in the current session providing more context into the the user's curent state of analysis, (2) a hybrid LSTM-RF classifier to ascertain if the RF model performs better when provided with more context captured by the LSTM in terms of the sequence of states that precede the current state of data analysis, and 
(3) a reinforcement learning-based DDQN~\cite{DoubleDQN} by modeling multi-class classification as an optimal $<$state,action$>$ prediction problem. Detailed experimental evaluation in Section~\ref{sec:evaluation} shows that the simpler RF classifier is highly efficient and provides better or comparable accuracy to the other models for $Intent_{BI}$ prediction.

\subsection{Step 2: Top-$k$ BI Pattern Prediction}
Traditional memory-based Collaborative Filtering (CF) algorithms exploit <user-item> similarity to make recommendations. We adapt the CF model for making BI pattern $P_{BI}$ prediction by modeling each session as a vector of states and compute session similarities between current and prior sessions to recommend the next $P_{BI}$. However, memory-based CF algorithms are not scalable as they tend to be computationally exhaustive and can potentially end up computing the similarities between the current state and the states among the entire set of prior user sessions. 
We explore two optimization techniques to address this issue. First, we designed and implemented an index-based CF approach $CF_{Index}$, a variant of the memory-based CF algorithms, for making the top-$k$ $P_{BI}$ predictions. We build a {\em Task Index} that prunes the space of prior user sessions whose states need to be compared with the current state to make the $P_{BI}$ prediction. 

Second, we implemented a matrix factorization-based CF model $CF_{SVD}$ that uses non-negative matrix factorization based on Singular Value Decomposition (SVD) for CF. This model serves as an approximation to the CF approach as it avoids the extensive state similarity computed by the latter over the entire list of prior user sessions. The $CF_{SVD}$ sorts the states based on the completed matrix scores and does not require the explicit computation of state similarities over all the states in the prior user sessions to recommend the top-$k$ BI patterns. 

An experimental evaluation of $CF_{Index}$ and $CF_{SVD}$ (Section~\ref{sec:evaluation}) for top-k BI Pattern prediction shows that although $CF_{SVD}$ is more efficient, it has a very low accuracy as compared to $CF_{Index}$. The $CF_{Index}$ approach has high accuracy and it is significantly more efficient than an exhaustive CF approach. We therefore use the $CF_{Index}$ approach for top-k $P_{BI}$ prediction. Next, we describe the $CF_{Index}$ approach in further detail. 

\subsubsection{Task Index}
\label{sec:task_index}
\begin{figure}[!ht]
    \begin{center}
    \includegraphics[width=\columnwidth]{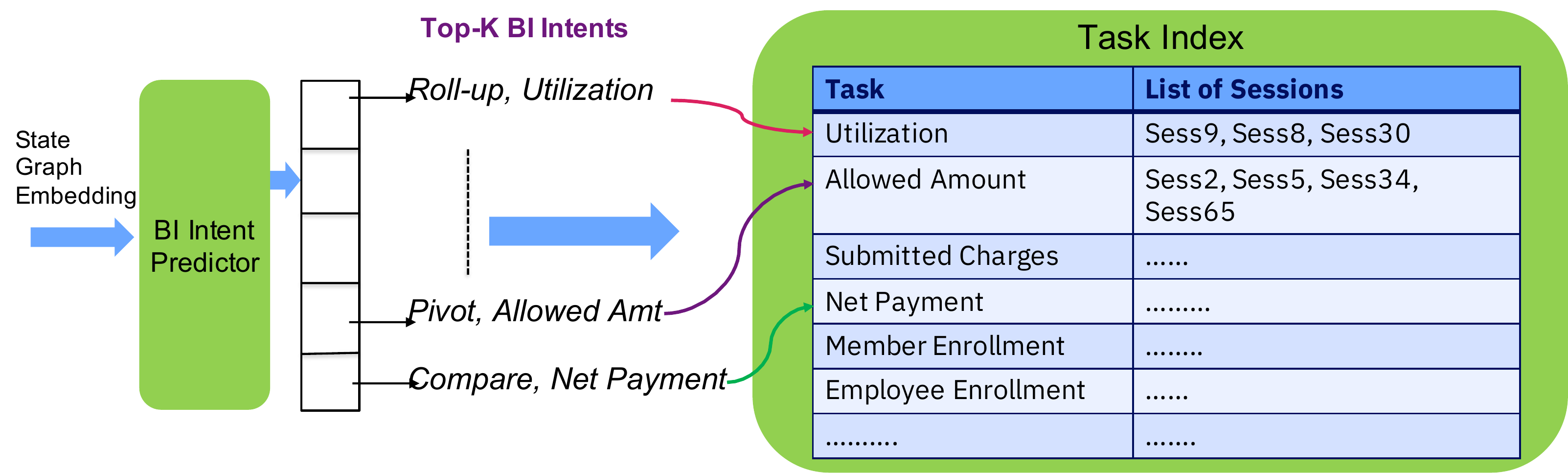} 
    %\vspace{-10pt}
    \caption{Finding relevant sessions using a task-based session index.}
    %\vspace{-15pt}
    \label{fig:sessionIndex}
    \end{center}
\end{figure}

The Task Index, is a task-based session index, that groups together sessions based on the $MG \in  Task_{US_k}$ (Figure~\ref{fig:sessionIndex}). For example, the session task for sessions 9, 8, and 30 contain the MG {\em Utilization} that analyzes the utilization of healthcare resources (e.g., hospitals and clinics) for admissions. The task index allows order one access to a list of sessions relevant for a particular MG. As seen in the figure, based on the top-$k$ $Intent_{BI}$s predicted by the $Intent_{BI}$ predictor, the system finds the relevant prior user sessions to make the $P_{BI}$ prediction using the task index, substantially pruning the search space of prior user sessions relevant for $P_{BI}$ prediction.

\subsubsection{$CF_{Index}$-based $P_{BI}$ Prediction}
\label{sec:indexBasedCollabFiltering}
\begin{figure}[!ht]
    \begin{center}
    \includegraphics[width=\columnwidth]{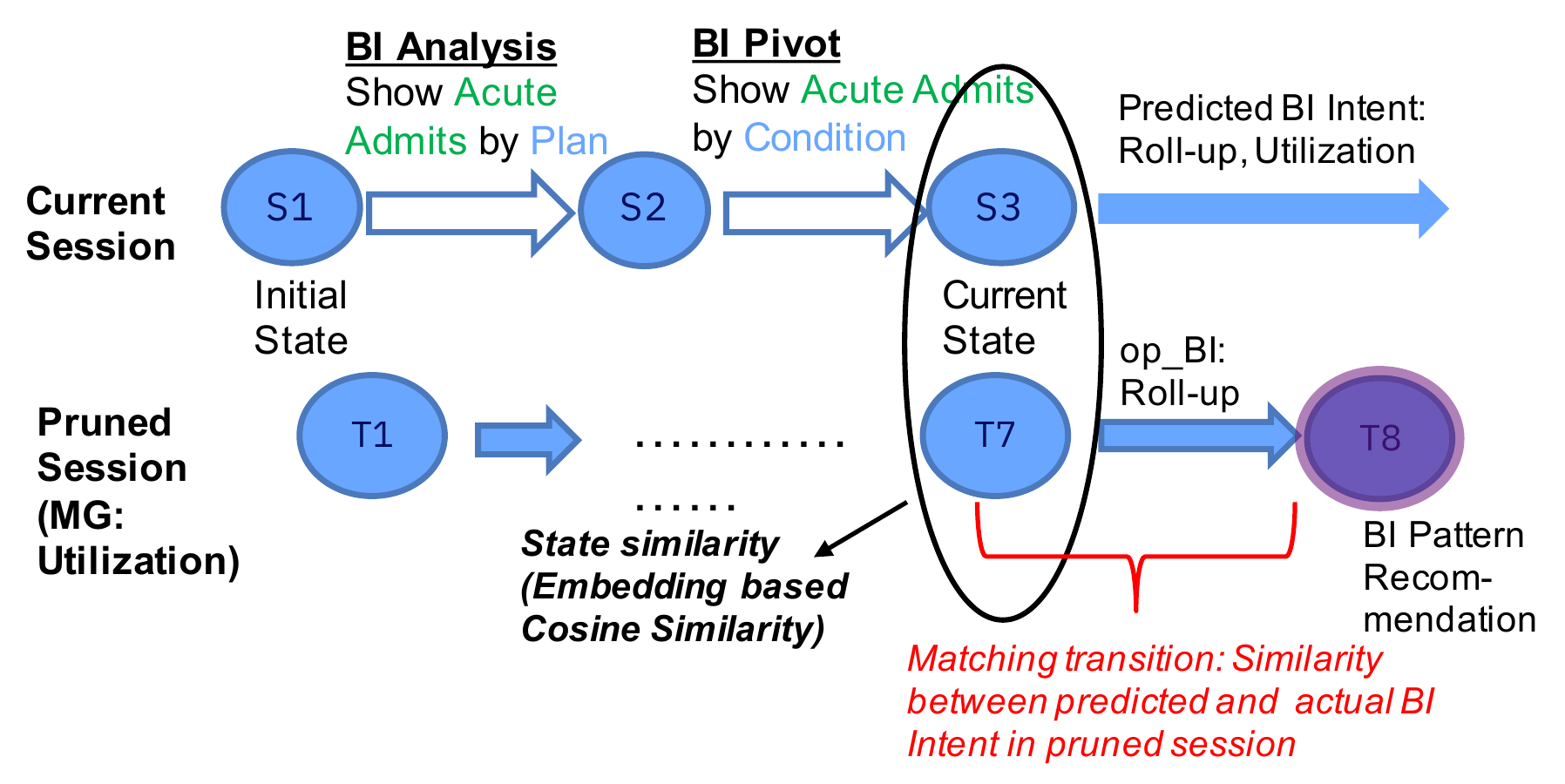} 
    %\vspace{-10pt}
    \caption{Index-based CF for BI query prediction.}
    %\vspace{-15pt}
    \label{fig:collabFilter}
    \end{center}
\end{figure}

Having identified the prior user sessions relevant to the MG predicted in the $Intent_{BI}$, the next step is to find the most similar state within these identified sessions to make the $P_{BI}$ prediction. Figure~\ref{fig:collabFilter} shows an example  current session with a current state $S_3$. The BI Intent predictor predicts <ROLL-UP, Utilization> as one of the top-$k$ $Intent_{BI}$. The system utilizes the task index (Figure~\ref{fig:sessionIndex}) to get a set of pruned sessions relevant to the MG {\em Utilization}.

The figure shows one example of a pruned session that contains Utilization as a MG in its session task. Within this session, the CF approach finds the most similar <state, $op_{BI}$> transition. The state $T_7$ is most similar to state $S_3$, the current state in the current session, and the $op_{BI}$ is ROLL-UP in the predicted BI Intent and the transition from state $T_7$ to $T_8$ in the pruned session. State $T_8$ is therefore now used to make the $P_{BI}$ prediction.
Equation~\ref{eqn:stateActionSim} provides a weighted function for computing the <state, $op_{BI}$> transition similarity based on the state and $op_{BI}$ similarities. We use the cosine similarity between the state graph embeddings $E_{S_i}$ of both these states to compute state similarity $Sim_{state}$. $Sim_{op_{BI}}$ is based on an exact match. In our current implementation, we set $w_s$ as 0.5 to provide equal weightage to state and $op_{BI}$ similarities, which has empirically been verified to provide the most accurate results.
\begin{equation}
Sim_{<state, op_{BI}>} = w_{s} * Sim_{state} + (1-w_{s}) * Sim_{op_{BI}}
\label{eqn:stateActionSim}
\end{equation}

\subsubsection{Refinement of BI Pattern Recommendations}
\label{sec:queryRefinement}
$P_{BI}$ refinement is the final step before making the top-$k$ $P_{BI}$ recommendations to the user. The motivation behind refinement is to utilize other interestingness metrics to enrich recommendations. We define interestingness of recommendations on the basis of how frequently prior users have queried a specific dimension along with a measure in the past. Users tend to project specific measures in conjunction with particular dimensions. Co-occurrence statistics record pairs of measures and dimensions along with their co-occurrence frequency in the prior workload of user sessions. Such co-occurrence frequency is a direct indicator of user interest in analysing certain measures along specific dimensions. Recommending a frequently co-occurring dimension in conjunction with a measure may result in a $P_{BI}$ that is highly likely to be accepted by the user. In the current implementation of \sys, we include query recommendation refinement based on co-occurrence statistics and leave the investigation of other methods as future work.

%% file: evaluation.tex
\section{Experimental Evaluation}
\label{sec:evaluation}

In this section, we provide a detailed evaluation of \sys. We first describe the dataset and workloads of BI patterns extracted from the conversational logs followed by the experimental setup and methodology, including evaluation metrics. We evaluate \sys\ components in terms of the effectiveness of
(1) our network representation learning models using GNNs to generate high quality compact state graph embeddings, and 
(2) our novel two-step approach for $Intent_{BI}$ prediction using a small amount of training data and efficient $P_{BI}$ prediction using our proposed $CF_{Index}$ approach. 
For an end-to-end system evaluation, we study the performance of \sys~on both real and synthetic workloads. We also compare the performance of \sys~with a state-of-the-art baseline~\cite{baseline} that predicts the entire $P_{BI}$ in one shot using an exhaustive CF approach. We compare the two systems in terms of prediction accuracy as well as latency with an underlying hypothesis that our two-step approach for $P_{BI}$ prediction would achieve accuracy comparable to the exhaustive baseline while providing substantial gain in prediction latency making it suitable for conversational BI systems.
Finally, we provide a user study that validates the quality and usefulness of our BI pattern recommendation for guided data analysis.

\subsection{Dataset and Workloads}
\label{sec:dataset}

\subsubsection{Datasets}

We use the Health Insights (HI) dataset (Section~\ref{sec:HealthInsights}) for our experimental evaluation. The BI ontology corresponding to this dataset contains 64 measures, 229 dimensions, 12 measure groups, and 13 dimension groups created by SMEs.
We also augment the ontology to create an Augmented Health Insights (AHI) dataset with 265 additional synthetic measures and 48 additional measure groups to enable studying the effect of different distributions of sessions per {\em Measure Group} (Session Task) on \sys's prediction accuracy and latency. 

\subsubsection{Workloads}

For the purpose of evaluation, we have used one real and five synthetic workloads against the HI dataset. These workloads consist of a set of user sessions, each containing a sequence of states extracted from the conversational logs against the HI dataset. We briefly describe each of these workloads below.

\textbf{Health Insights workload (HIW).} HIW is a real workload collected from the logs of conversational interaction between our conversational BI system~\cite{DBLP:journals/pvldb/QuamarOMMNK20} and a mix of technical and non-technical business users. The user interactions were recorded as sessions wherein each user was assigned 5 different tasks in terms of MGs such as utilization of healthcare resources (Utilization), cost incurred by insurance (Net Payment), etc. Each user session consisted of multiple turns of conversation with users issuing separate analysis queries. The responses to user queries were displayed as charts and the users terminated a session when the assigned task was complete. We collected a total of 125 user sessions across different session tasks over the course of one month wherein the average user session length ranged between 5 to 8 queries issued by the user to accomplish the assigned tasks.

\begin{table}
\small
\caption{Synthetic workloads based on the distribution of ${op}_{BI}$ transition probability.}
\begin{tabular}{ |p{1.4cm}||p{1.5cm}|p{2.2cm}| }
\hline
Workload & Distribution & Parameters\\
\hline
BT-Exp & Exponential & mean=0.5\\
BT-Gamma & Gamma & shape =1, scale=1\\
BT-Uniform & Uniform & value $\in$ [0.0, 1.0]\\
BT-Normal & Normal & mean=0, stddev=1 \\
\hline
\end{tabular}
\label{tab:workloads_transition}
\end{table}

\begin{table}
\caption{Synthetic workloads based on the distribution of \# user sessions per task.}
\begin{tabular}{|c||c|p{1.9cm}|p{1.8cm}|}
\hline
Workload & Distribution & [Min,Max] \# session per task (HI) & [Min,Max] \# session per task (AHI)\\
\hline
ST-Exp & Exponential & [3,20] & [1,8] \\
ST-Gamma & Gamma & [3,27] & [1,9]\\
ST-Uniform & Uniform & [10,11] & [2,3]\\
ST-Normal & Normal & [3,13] & [1,5]\\
\hline
\end{tabular}
\label{tab:workloads_user_session}
%\vspace{-20pt}
\end{table}

\textbf{Synthetic workloads} are generated against the HI dataset as well. We validate our system performance upon a variety of synthetic workloads that broadly differ in terms of (1) the distribution of the transition probabilities between different ${op}_{BI_s}$ observed in a user data analysis session and (2) the distribution of MGs (e.g., {\em Utilization}, {\em Net Payment}, etc.) in the session tasks of user sessions. For example, a uniform distribution would evenly distribute the number of user sessions containing a particular MG in their session task across different MGs, as opposed to an exponential distribution wherein a few MGs would have a much higher number of user sessions compared to others. Tables~\ref{tab:workloads_transition} and~\ref{tab:workloads_user_session} provide the statistics of the synthetic workloads used in our experimental evaluation. 

\subsection{Experimental Setup and Methodology}

\subsubsection{Settings and Configuration}

We conducted our experiments on a machine with 2.3 GHz 8-Core Intel Core i9 processor and 64 GB 2667 MHz DDR4 RAM running Mac OS. We implemented the end-to-end system using Python 3.7.8. We used PyTorch as the deep learning platform with different libraries for the implementation of GNNs~\cite{DiffPool-DGL} and LSTMs~\cite{LSTMs-PyTorch}. We used scikit-learn for implementing random forests~\cite{RF-scikit} and SVD-based CF~\cite{CF-SVD}. We have implemented index-based CF and value networks for Deep Q-Learning while using PyTorch~\cite{DQN-PyTorch} to create and train the neural net.

\subsubsection{Evaluation Metrics and Methodology}

We used 5-fold cross-validation to experimentally evaluate the different components of \sys.
We used 662 queries across 100 training sessions and 140 queries across 25 test sessions in all the folds. 
For the evaluation of state representation model, we report F1-score, root mean square error (RMSE), and accuracy. 
For the evaluation of $Intent_{BI}$ and $P_{BI}$ predictions, we report the Jaccard similarity between the expected (${P}_{BI,exp}$) and predicted (${P}_{BI,pred}$) BI patterns, respectively. 
\begin{equation}
JaccSim(P_{BI,exp}, P_{BI,pred}) = \frac{|P_{BI,exp} \cap P_{BI,pred}|}{|P_{BI,exp} \cup P_{BI,pred}|}
\label{eqn:JaccSim}
\end{equation}

\subsubsection{Baseline}
\label{sec:baseline}

We compare the performance of \sys~ in terms of latency and accuracy with a session summary-based CF baseline~\cite{baseline} that was originally developed for SQL query recommendation. We adapt the baseline for $P_{BI}$ recommendation. We chose ~\cite{baseline}  over a na\"ive CF approach that would exhaustively scan all the states within the prior user sessions in order to find the most similar state to the current state.~\citet{baseline} avoid this exhaustive search by computing a session summary for each prior user session and use this summary to prune the set of sessions not relevant to the current session.

\subsection{\sys~ System Component Evaluation}

We divide the evaluation of \sys's components into three parts (1) state representation, (2) $Intent_{BI}$ prediction, and (3) BI pattern $P_{BI}$ recommendation, as described next.

\subsubsection{Evaluation of State Representation}

\label{sec:stateEval}
\begin{figure*}[tb]
	\centering
	\captionsetup[subfigure]{justification=centering}
	\begin{subfigure}[t]{0.24\textwidth}
		\centering
		\includegraphics[width=\linewidth]{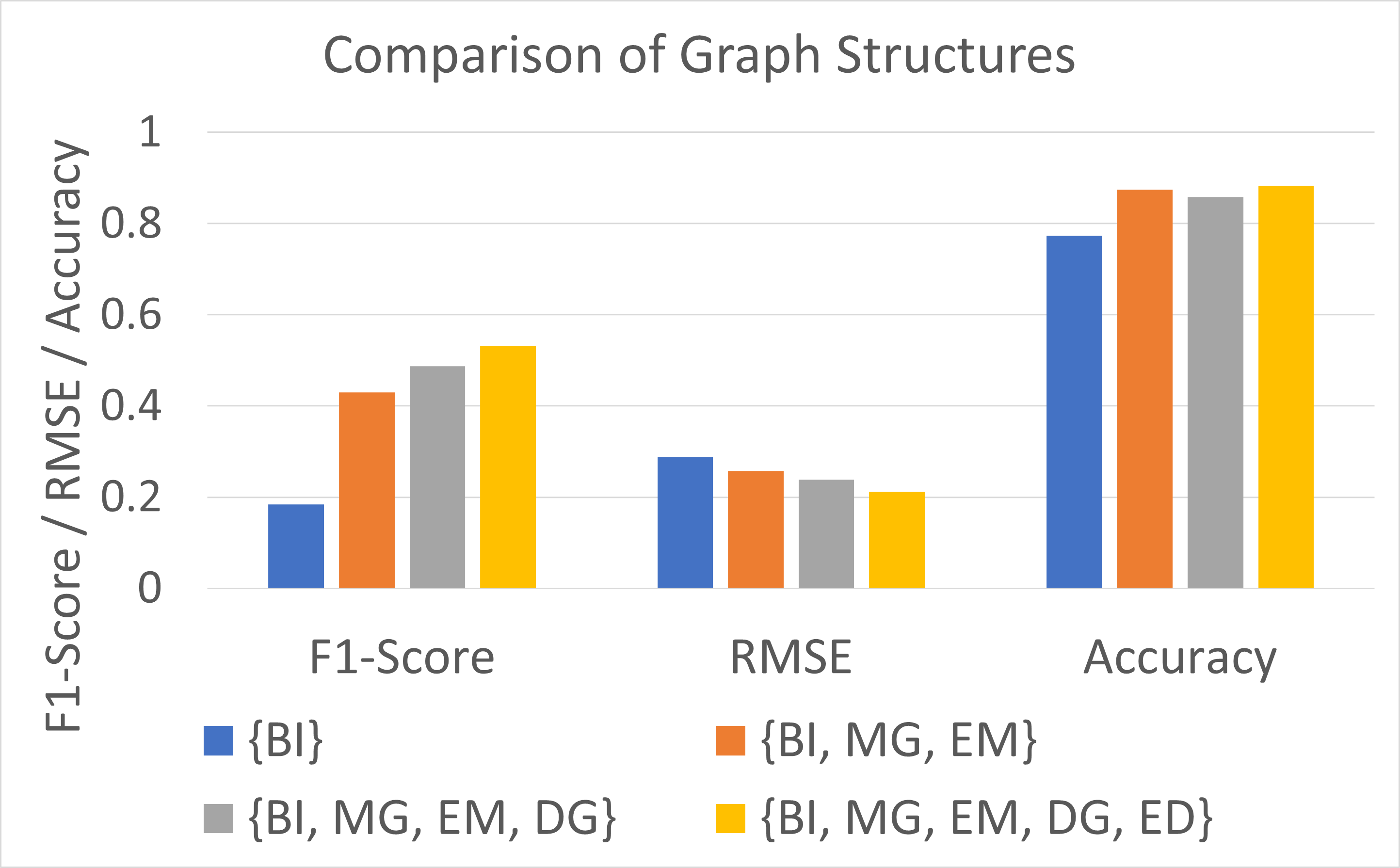}
		\caption{Embedding quality vs \\level of enrichment}
		\label{fig:embedQuality}
	\end{subfigure}
	\begin{subfigure}[t]{0.24\textwidth}
		\centering
		\includegraphics[width=\linewidth]{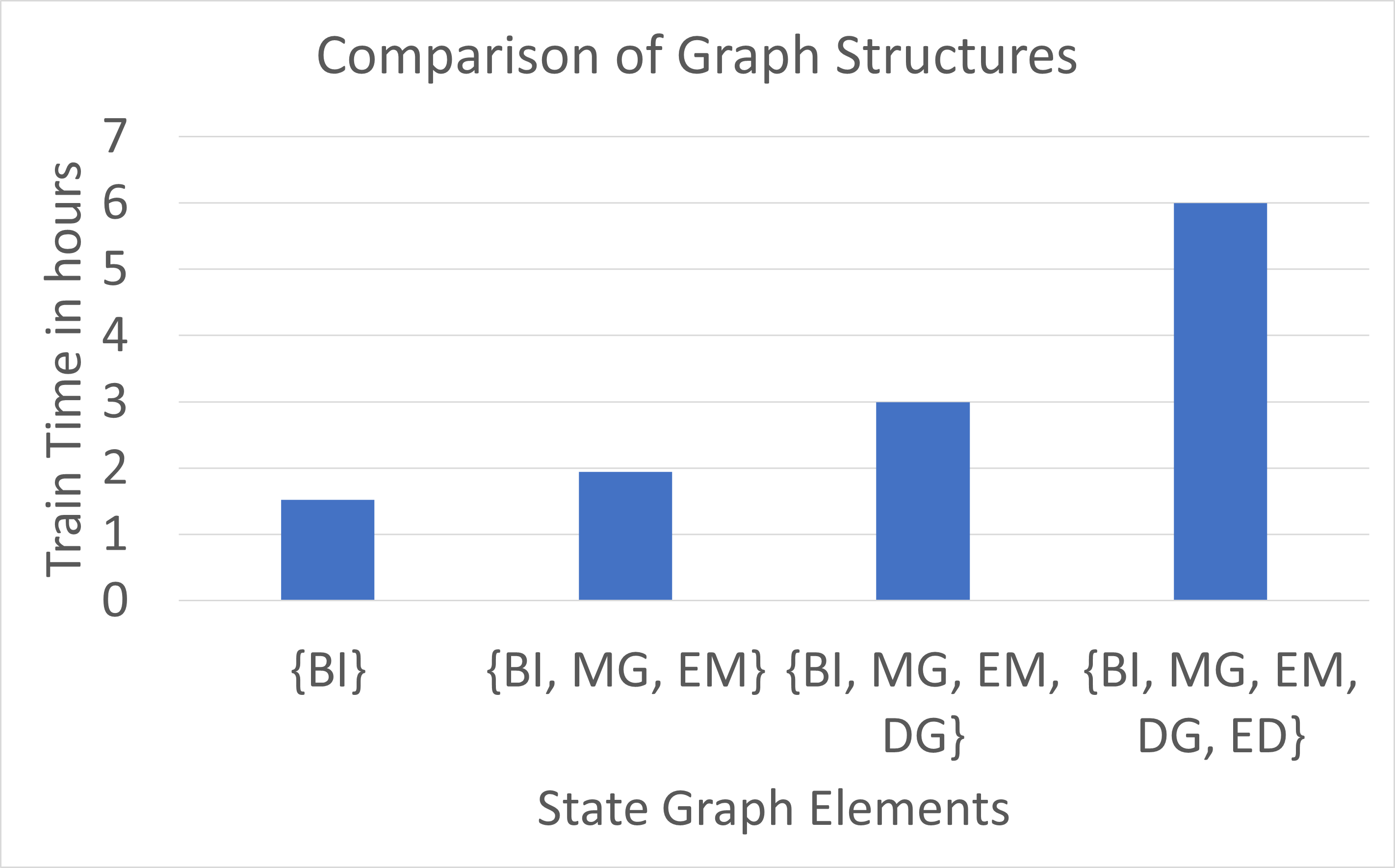}
		\caption{Model training time vs \\level of enrichment}
		\label{fig:embedLatency}
	\end{subfigure}
	\begin{subfigure}[t]{0.24\textwidth}
		\centering
		\includegraphics[width=\linewidth]{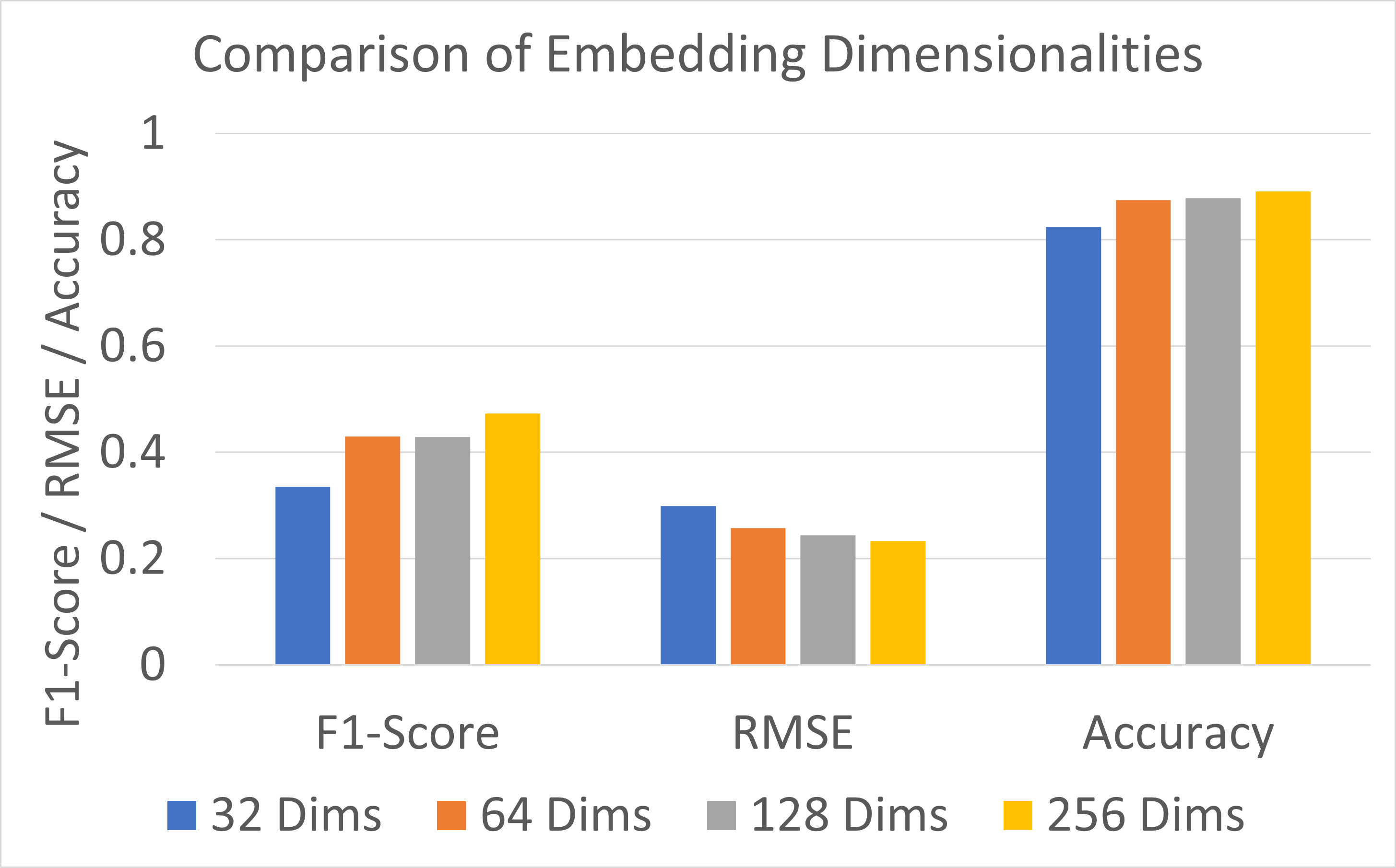}
		\caption{Embedding quality vs \\$\#$Dimensions}
		\label{fig:embedQualityDims}
	\end{subfigure}
	\begin{subfigure}[t]{0.25\textwidth}
		\centering
		\includegraphics[width=\linewidth]{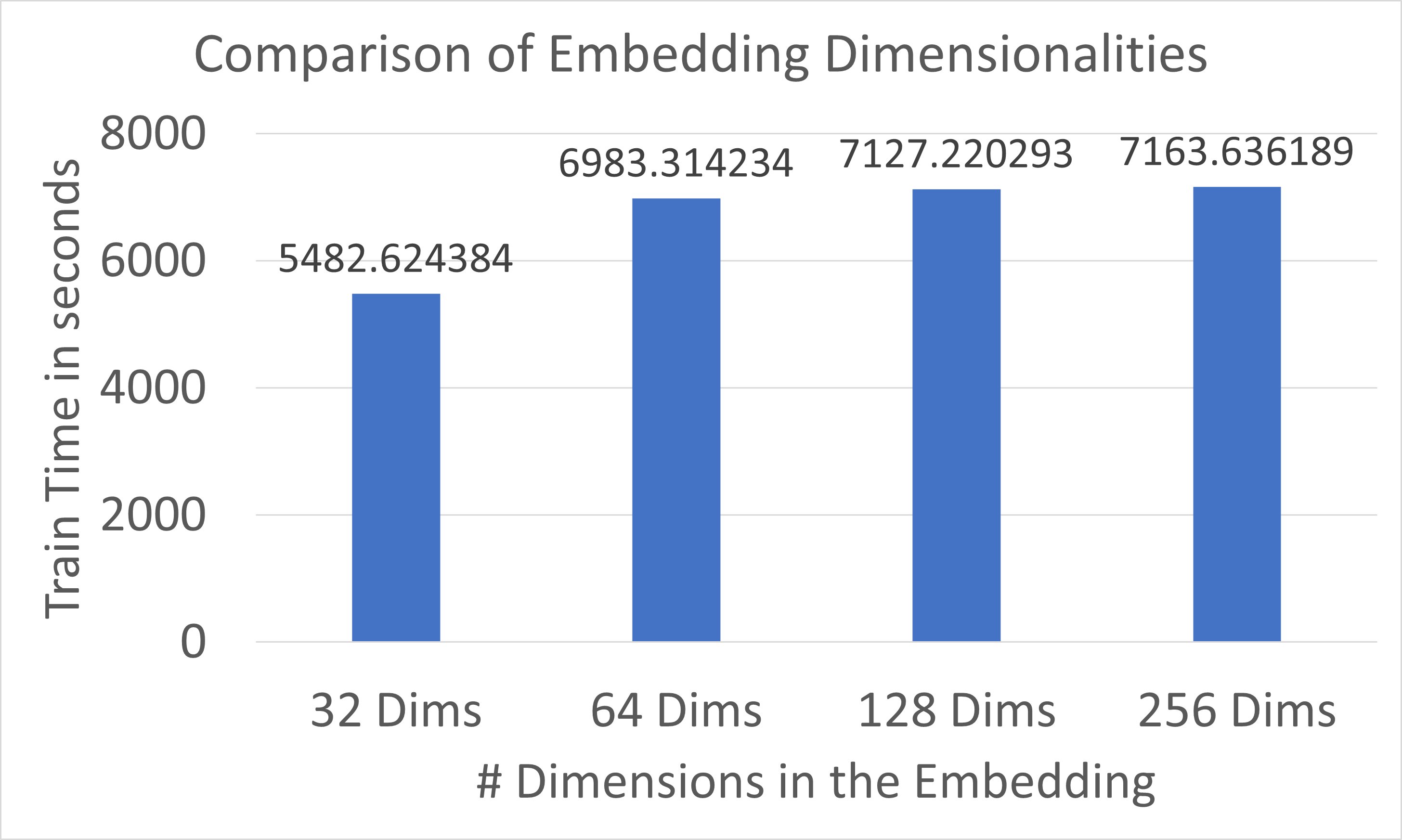}
		\caption{Model training time vs \\$\#$Dimensions}
		\label{fig:embedLatencyDims}
	\end{subfigure}
	%\vspace*{-.4cm}
	\caption{Evaluation of state representation.}
	%\vspace*{-.4cm}
	\label{fig:stateRepEval}
\end{figure*}

We evaluate the state graph embeddings in terms of accuracy and training time of the GNN model by (1) varying the levels of enrichment of the state graph with elements from the ontology neighborhood $ON$ and (2) varying the number of embedding dimensions.

Figure~\ref{fig:stateRepEval} shows the 5-fold evaluation results of state representation using GraphSAGE~\cite{NIPS2017_5dd9db5e}. The training and test sets consist of sampled matching and non-matching pairs of states created by using a similarity (Equation~\ref{eqn:JaccardSimilarity}) threshold. State pairs with $Sim > 0.5$ are considered as matching, and the rest are considered as non-matching. Evaluation on the test set checks whether the latent space similarity upon the pair of state graph embeddings exceeds 0.5, to determine the label as ``matching'' or ``non-matching''. While we also present F1-score and RMSE, the accuracy metric is more relevant as unlike the F1-score, it gives equal weightage to detecting both matching and non-matching pairs of states. 

Figure~\ref{fig:embedQuality} shows the variation of embedding quality (for 64-dimensional embeddings) with different levels of enrichment: 
(1) $\{BI\}$ – No enrichment; only the root node and the elements of $P_{BI}$ are included in the state graph.
(2)	$\{BI, MG, EM\}$ – includes the elements in $P_{BI}$ along with the measure groups (MG) and expanded measures (EM) from the $ON$. 
(3) $\{BI, MG, EM, DG\}$ –  includes the elements in the $P_{BI}$, MG, EM along with the dimension groups (DG) from the $ON$. 
(4) $\{BI, MG, EM, DG, ED\}$ – includes the expanded dimensions (ED) from the $ON$ along with the information included in ${P}_{BI}$, MG, EM, DG.

The results in Figure~\ref{fig:embedQuality} indicate that as we increase the level of enrichment of the state graphs with information from the $ON$, the embedding quality both in terms of F1-score and accuracy increases, achieving an overall accuracy of 0.88 for state graphs enriched with \{BI, MG, EM, DG, ED\}. The reason is that only relying on the queried elements to represent a state graph leads to a rigid similarity criterion, as it requires the queries in a state graph pair to contain the exact same elements for them to be similar. On the other hand, enriching the state graph with rich semantic information from  $ON$ relaxes the similarity criterion to include semantically similar state graphs that might not have been seen in prior workloads, thereby allowing us to make meaningful recommendations. 

Figure~\ref{fig:embedLatency} shows the time required to train the GNN model to generate the different graph embeddings. We see that training time for \{BI,MG,EM\} is almost 2 orders of magnitude lower than \{BI,MG,EM,DG,ED\}, whereas both the graphs have comparable accuracies (0.87 vs. 0.88). Hence, for the remainder of the experiments, we use \{BI,MG,EM\} as the state graph representation. 

Figures~\ref{fig:embedQualityDims} and~\ref{fig:embedLatencyDims} show the variation of embedding quality and model training time, respectively, as the length of the embedding vector ranges from 32 to 256. We notice a significant increase in accuracy (from 0.82 to 0.87) between 32 and 64 dimensions while there is no noticeable increase in accuracy beyond 64 dimensions. Therefore, for the sake of compactness, we use 64 dimensional embeddings for state representation in all the remaining experiments. We would like to note here that although there is an increase in the model training time between 32 and 64 dimensions, we train the embedding model in an offline pre-processing step and hence do not consider the difference to be critical.

\begin{comment}
  \begin{figure}[!ht]
	\begin{center}
		\includegraphics[scale=0.7]{figures/embedResults.png} 
		\vspace{-10pt}
		\caption{Embedding Accuracy vs. Graph Expansion}
		\vspace{-15pt}
		\label{fig:embedResults}
	\end{center}
\end{figure}
\end{comment}

\subsubsection{Evaluation of top-$k$ BI Intent Prediction}

In this section, we evaluate the accuracy of the top-$k$ $Intent_{BI}$ prediction, which is the first step in our two-step approach for $P_{BI}$ prediction.
As mentioned in Section~\ref{sec:BIIntentPrediction}, we trained and tested several different multi-class classifier models for top-$k$ $Intent_{BI}$ prediction including Random Forests (RFs), LSTMs, a hybrid RF+LSTM model, as well as a Reinforcement Learning based Double DQN model~\cite{DoubleDQN}. Reinforcement Learning can model the prediction task as the selection of an optimal action ($op_{BI}$) given the state information, using appropriate reward functions to quantify the conditional effectiveness of various actions under a given state. We model Double DQNs as discriminative, supervised learners for $Intent_{BI}$ prediction by setting the reward function in such a way that the selection of $Intent_{BI}$ based on the next $op_{BI}$ from the  workload of prior sessions fetches the highest reward at a given state during training. The objective function was set to minimize the loss between the Q-value from the Bellman equation update and the Q-value predicted by the network.

\begin{figure*}[tb]
	\centering
    \begin{subfigure}[t]{0.36\textwidth}
		\centering
		\includegraphics[width=\linewidth]{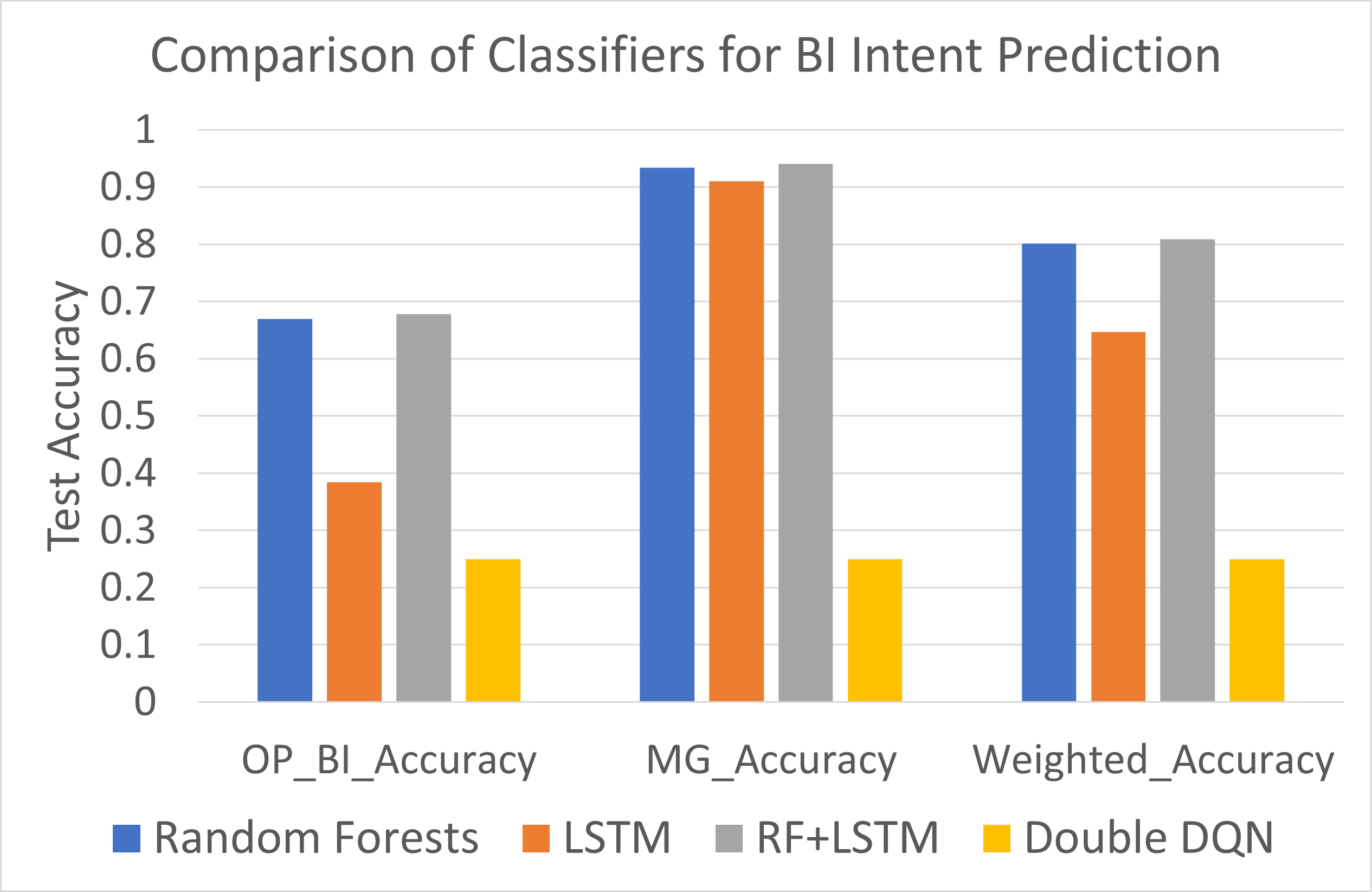}
		\caption{Comparison of classifiers}
		\label{fig:intentPred-1}
	\end{subfigure}
	\begin{subfigure}[t]{0.31\textwidth}
		\centering
		\includegraphics[width=\linewidth]{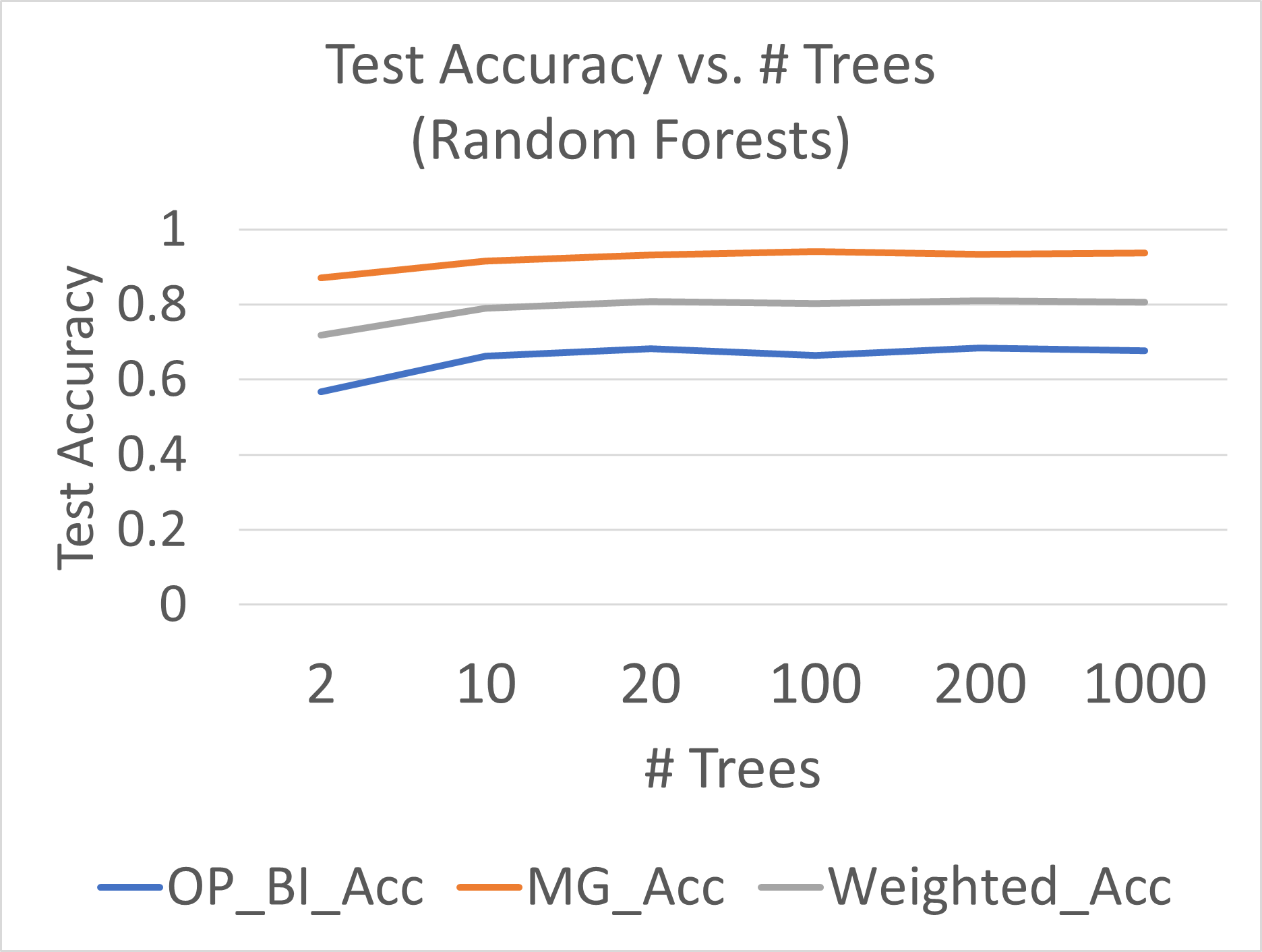}
		\caption{Varying $\#$ trees}
		\label{fig:intentPred-2}
	\end{subfigure}
	\begin{subfigure}[t]{0.32\textwidth}
		\centering
		\includegraphics[width=\linewidth]{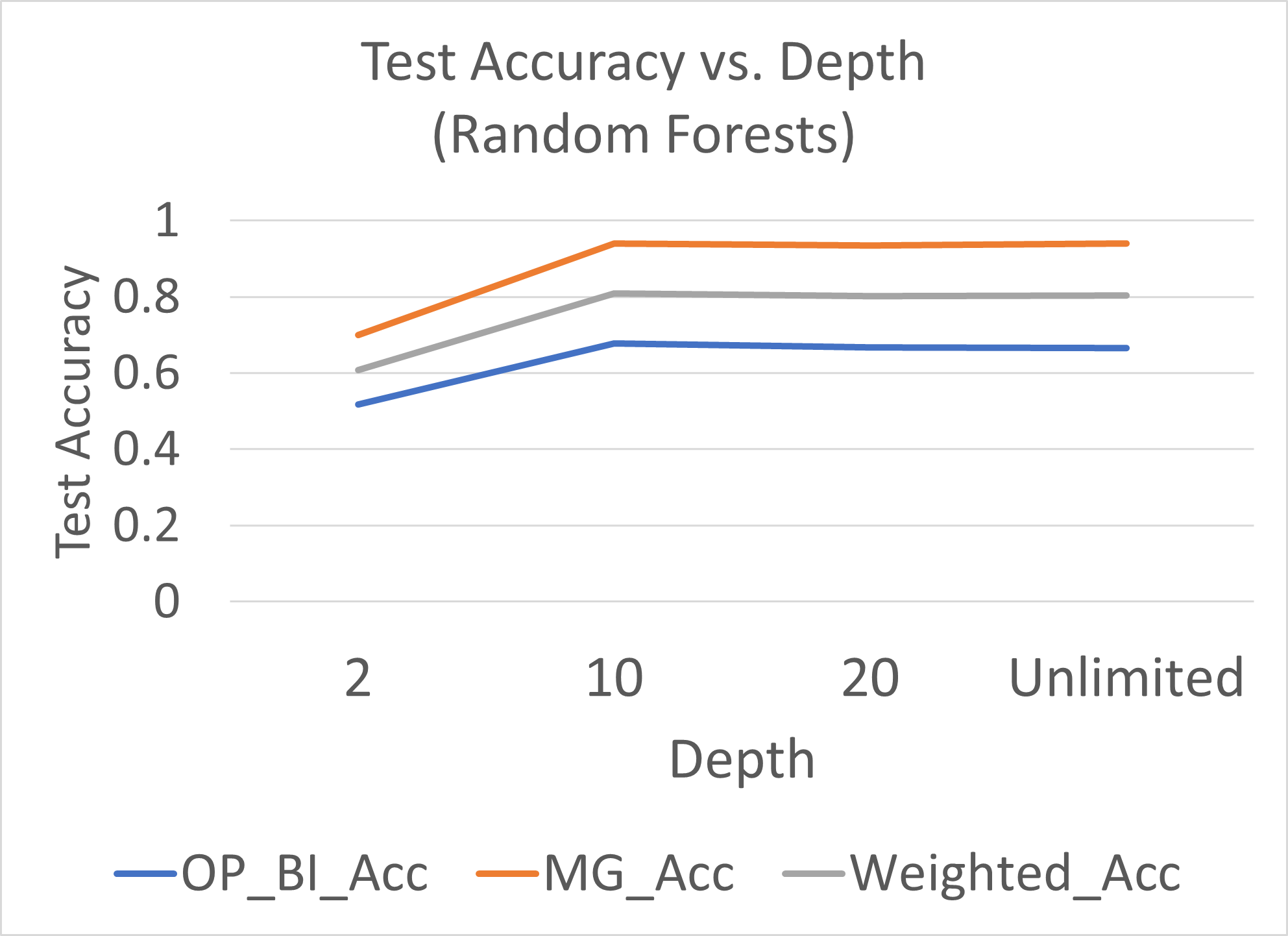}
		\caption{Varying $\#$ depth}
		\label{fig:intentPred-3}
	\end{subfigure}
%\vspace*{-.4cm}
	\caption{Evaluation of ML models for BI intent prediction.}
%\vspace*{-.4cm}
	\label{fig:intentPredResults}
\end{figure*}

Figure~\ref{fig:intentPredResults} shows the results for the top-$k$ $Intent_{BI}$ predictors. The accuracy of top-$k$ $Intent_{BI}$ prediction was measured by comparing the expected $Intent_{BI}$ obtained from the next state in the workload of prior user sessions, against the predicted $Intent_{BI}$. We measured the test accuracy from 5-fold cross validation as a weighted combination of ${op}_{BI}$ accuracy and $Task_{US_k}$ accuracy giving equal weights (0.5) to predicting both the elements of $Intent_{BI}$. 
As shown in Figure~\ref{fig:intentPredResults}, RF performs very well compared to other models. We observe that feeding sequences of states (BI patterns) does not bring a significant benefit based on the evaluation results on the LSTM and hybrid RF + LSTM models. This can be explained by our empirical observation that, learning individual state transitions well is equivalent to learning the sequences effectively. 
Figures~\ref{fig:intentPred-2} and~\ref{fig:intentPred-3} show the variation of RF accuracy with the number of trees and the depth of trees as a part of an ablation study. We notice that among the various settings for $\#$ trees in the ensemble and depth, using as few as 20 trees with a depth of 10 provides competitive performance as compared to 100 trees and unlimited depth. 

The evaluation results highlight the effectiveness of our two step approach which limits the search space of $Intent_{BI}$ to $\mathcal{S}_{{Intent}_{BI}}$ (Equation~\ref{eqn:search_space_BI_Intent}). This allows us to train simpler multi-class classifiers such as RF to achieve high accuracy using a small amount of training data. On the other hand, the poor performance of Double DQNs can be explained by the fact that DQNs serve as approximations to the Q-table and thereby require a significant amount of training data before producing robust and convergent predictions that align well with a materialized in-memory Q-table.

\subsubsection{Evaluation of Top-$k$ BI Pattern Prediction}

\begin{figure*}[htb]
		\centering
		\begin{subfigure}[t]{0.31\textwidth}
			\centering
			\includegraphics[width=\linewidth]{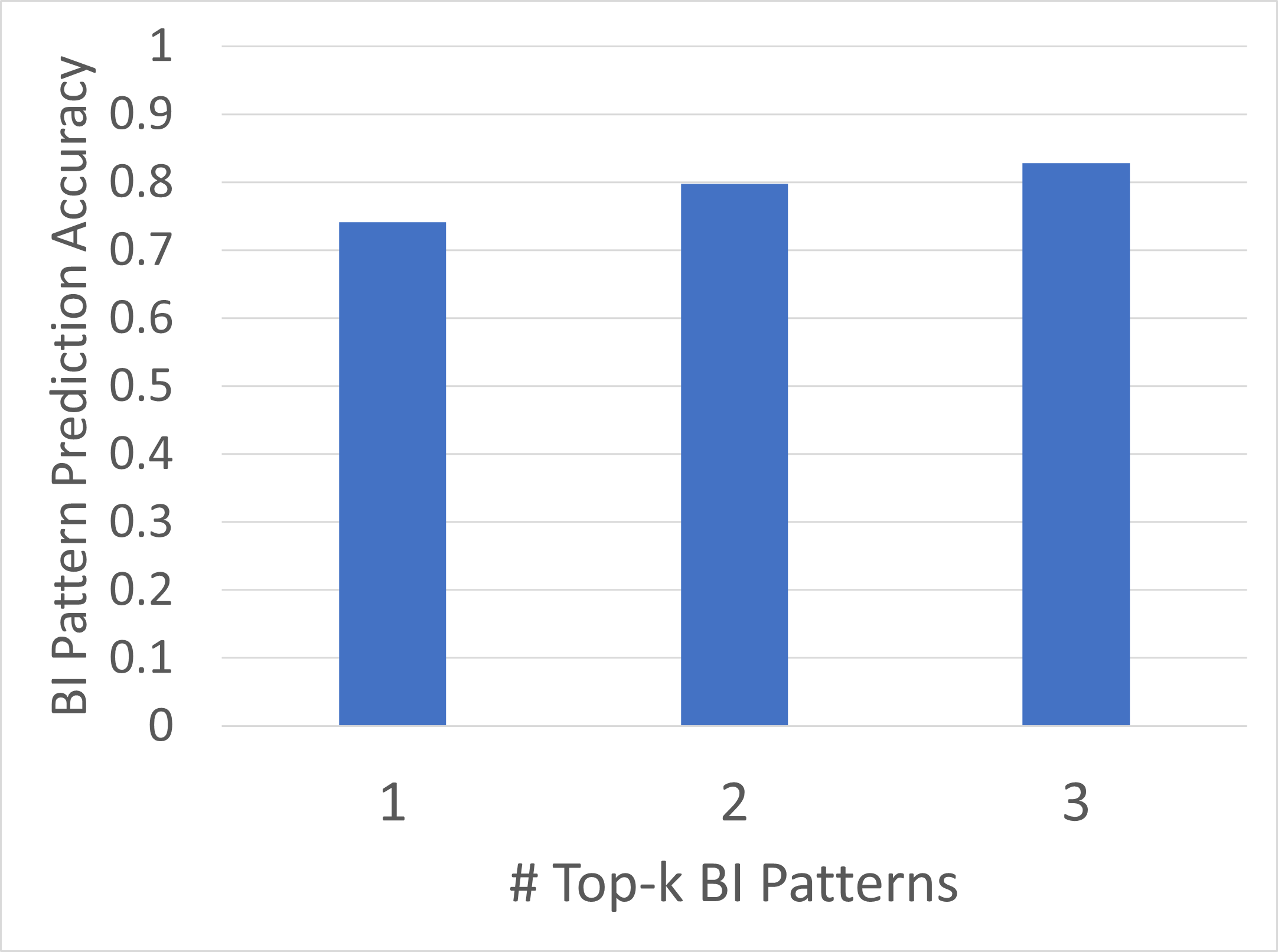}
			\caption{Evaluation of Index-based CF}
			\label{fig:CFQuality}
		\end{subfigure}
		\begin{subfigure}[t]{0.34\textwidth}
			\centering
			\includegraphics[width=\linewidth]{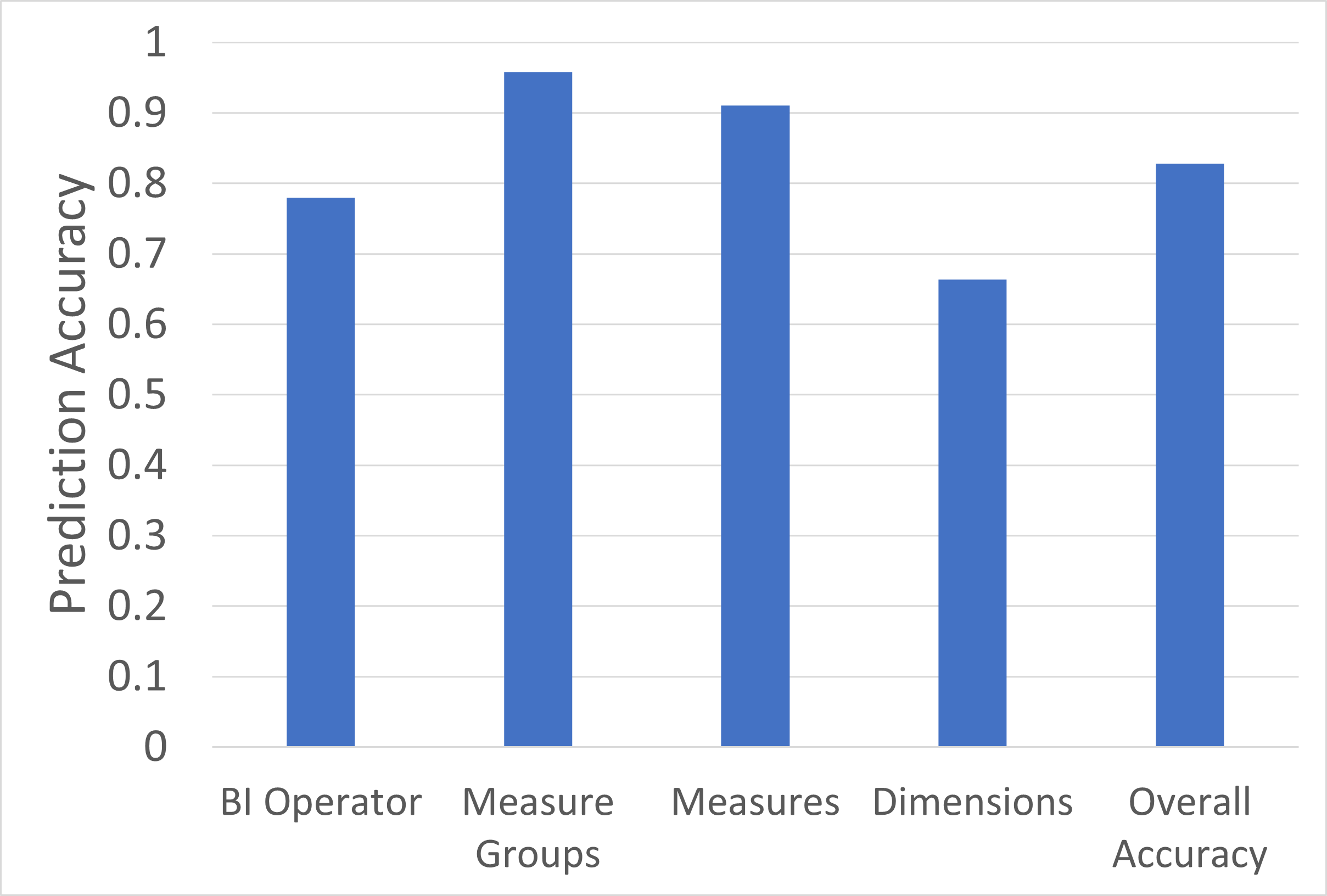}
			\caption{Quality breakdown}
			\label{fig:CFQualityBreakdown}
		\end{subfigure}
        \begin{subfigure}[t]{0.33\textwidth}
            \includegraphics[width=\linewidth]{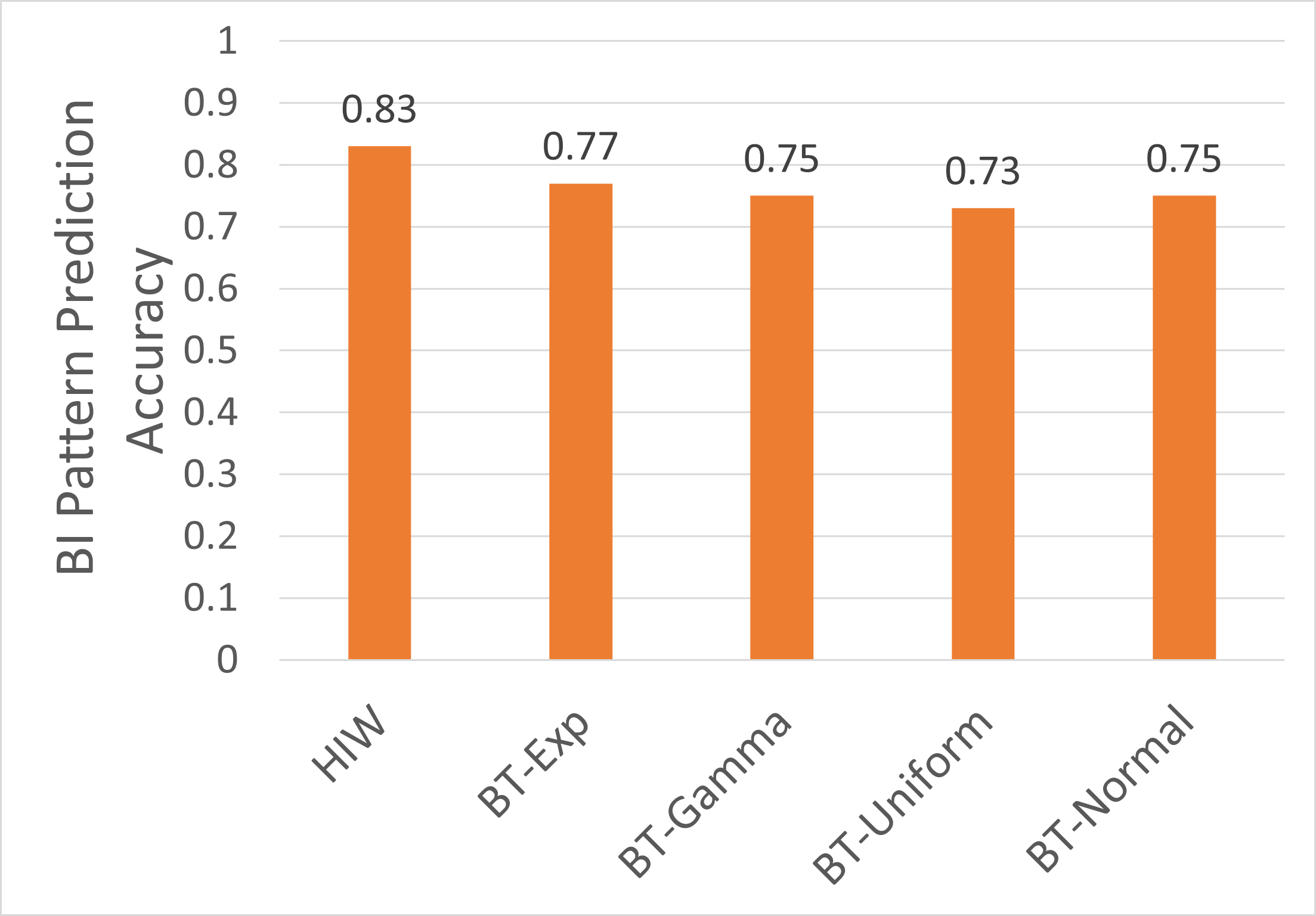} 
    		\caption{Effect of BI pattern transition probability distributions.}
    		\label{fig:statDist}
        \end{subfigure}
		\caption{Evaluation of Top-$k$ BI pattern prediction.}
		\label{fig:CFResults}
	\end{figure*}

In this section, we evaluate the accuracy of top-$k$ $P_{BI}$ prediction, the second step in our two-step approach. 
We evaluate our proposed index-based CF approach $CF_{Index}$ in terms of the overall prediction accuracy, and we provide a breakdown of this accuracy for different elements within the BI pattern, $P_{BI}$, such as $op_{BI}$, measures, dimensions and the session task represented by the measure group. We choose the top-$k$ $Intent_{BI}s$ predicted by RF multi-class classifier model with $k$ ranging from 1 to 3, and recommend one BI pattern, $P_{BI}$, per chosen $Intent_{BI}$ (Ref Figure~\ref{fig:CFQuality}). We set the upper bound of $k$ to be 3, in line with earlier works on SQL query prediction (e.g.,~\cite{DBLP:journals/tods/MeduriCS21}) which recommend top-3 queries at most, taking the cognitive ability of real human users into account.
We notice that increasing the value of $k$ from 1 to 3 results in higher accuracy, as it increases the likelihood for the top-$k$ recommendations to contain the expected BI pattern from the actual next query. We measure the  %overall accuracy as the 
Jaccard similarity between the predicted and the actual next BI pattern from the workload using Equation~\ref{eqn:JaccSim}. 

Figure~\ref{fig:CFQualityBreakdown} shows the breakdown of prediction accuracy with respect to different elements within the predicted BI pattern $P_{BI}$ that is the closest to the expected $P_{BI}$, among the top-3 predicted candidates. We see that prediction accuracy for dimensions is a bit lower than the prediction accuracy for the rest of the queried elements which could be attributed to the much larger number of dimensions as compared to measures and measure groups in the underlying dataset. In order to improve the accuracy of the predicted dimensions, we refine the final query recommendations by exploiting the co-occurrence statistics between the measures and dimensions which co-occur within a BI pattern, as discussed in Section~\ref{sec:queryRefinement}. \sys~recommends dimensions that occur most frequently with the measures in the predicted BI pattern. Table~\ref{tab:infDim} shows the improvement in dimension prediction accuracy by using this technique for up to 3 such dimensions as compared to the default implementation with 0 inferred dimensions not exploiting the co-occurrence statistics.

As mentioned in Section~\ref{sec:indexBasedCollabFiltering}, we compare our index-based CF approach, $CF_{Index}$, against an approximate approach, $CF_{SVD}$.
While $CF_{Index}$ achieves a maximum accuracy of 0.83 for top-3 BI pattern recommendation, $CF_{SVD}$ yields a maximum accuracy of 0.39  for the same setting. 
We varied the number of latent dimensions from 5 to 100, and found the best accuracy at 80 latent dimensions.
The total prediction time incurred by $CF_{SVD}$ is 0.02 sec as compared to 0.2 sec incurred by $CF_{Index}$. However, the low accuracy of $CF_{SVD}$ shows that the approximation is weak and matrix scores alone cannot be used to recommend the next BI query. This is because, a BI pattern with the highest score cannot simply be recommended as next state without knowing the semantic closeness between the current BI pattern and the recommended BI pattern. In the remainder of the experiments, we use $CF_{Index}$ for BI pattern recommendation as it yields a much higher accuracy. 

\begin{table}
\caption{Effect of co-occurrence statistics to improve dimension prediction accuracy.}
\begin{tabular}{ |c||c|c|c|c|}
\hline
$\#$\textbf{Inferred Dimensions} & 0 & 1 & 2 & 3 \\
\hline
\textbf{Dimension Accuracy} & 0.6635 & 0.6858 & 0.7018 & 0.7256\\
\hline
\end{tabular}
\label{tab:infDim}
\end{table}

\subsection{End-to-end System Evaluation}
We evaluate the end-to-end system performance of \sys~using different real and synthetic workloads, and compare its performance to a baseline system. 
In all the experiments, we set $k$ to 3 when predicting the top-$k$ BI patterns, taking the cognitive ability of the human user into consideration~\cite{DBLP:journals/tods/MeduriCS21}.

\subsubsection{\sys~end-to-end performance on different workloads}
Figure~\ref{fig:statDist} shows the end-to-end performance of \sys~in terms of accuracy for top-$k$ BI pattern prediction for the real HIW workload and four different synthetic workloads that differ in the transition probabilities between the different ${op}_{BI}$s between successive states in a user session. The high prediction accuracies across different distributions highlight that \sys~is robust in predicting the next BI pattern %accurately 
regardless of the underlying transition distributions and can adapt to different workloads.
We also notice that 
the accuracy of top-$k$ BI pattern prediction for the HIW workload (0.83) is higher than the best accuracy recorded on the synthetic counterpart (0.77). The reason for this is the inherent skew in the HIW workload that was created from a user study, wherein users were assigned a fixed set of MGs to investigate, and hence were biased towards using a subset of the BI operations more prominently than the others. 

\subsubsection{Exhaustive CF Baseline Comparison}

\begin{figure*}[tb]
	\centering
	\begin{subfigure}[t]{0.23\textwidth}
		\centering
		\includegraphics[width=\linewidth]{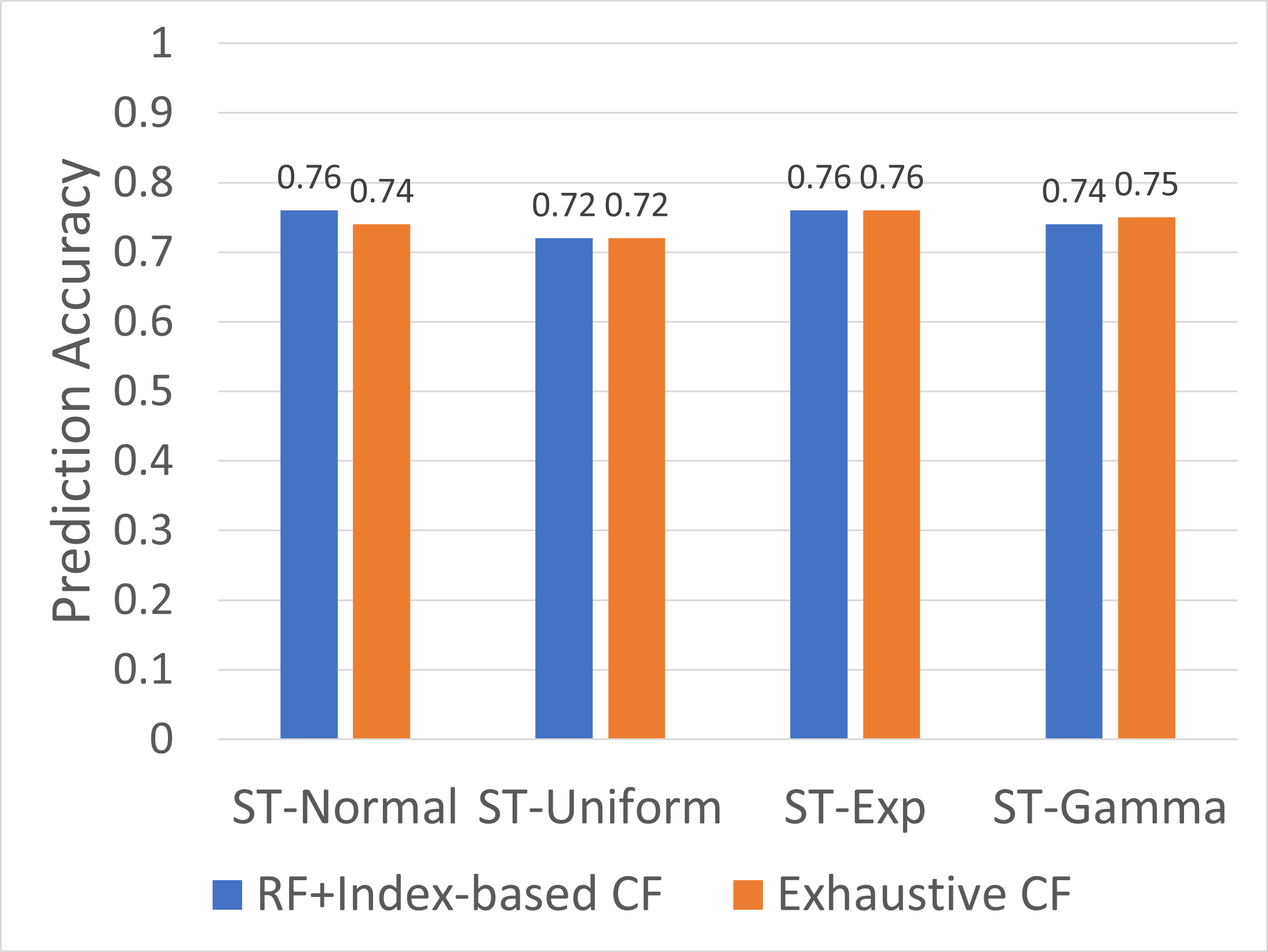}
		\caption{Prediction accuracy}
		\label{fig:CFQualityMGSynth}
	\end{subfigure}
	\begin{subfigure}[t]{0.24\textwidth}
		\centering
		\includegraphics[width=\linewidth]{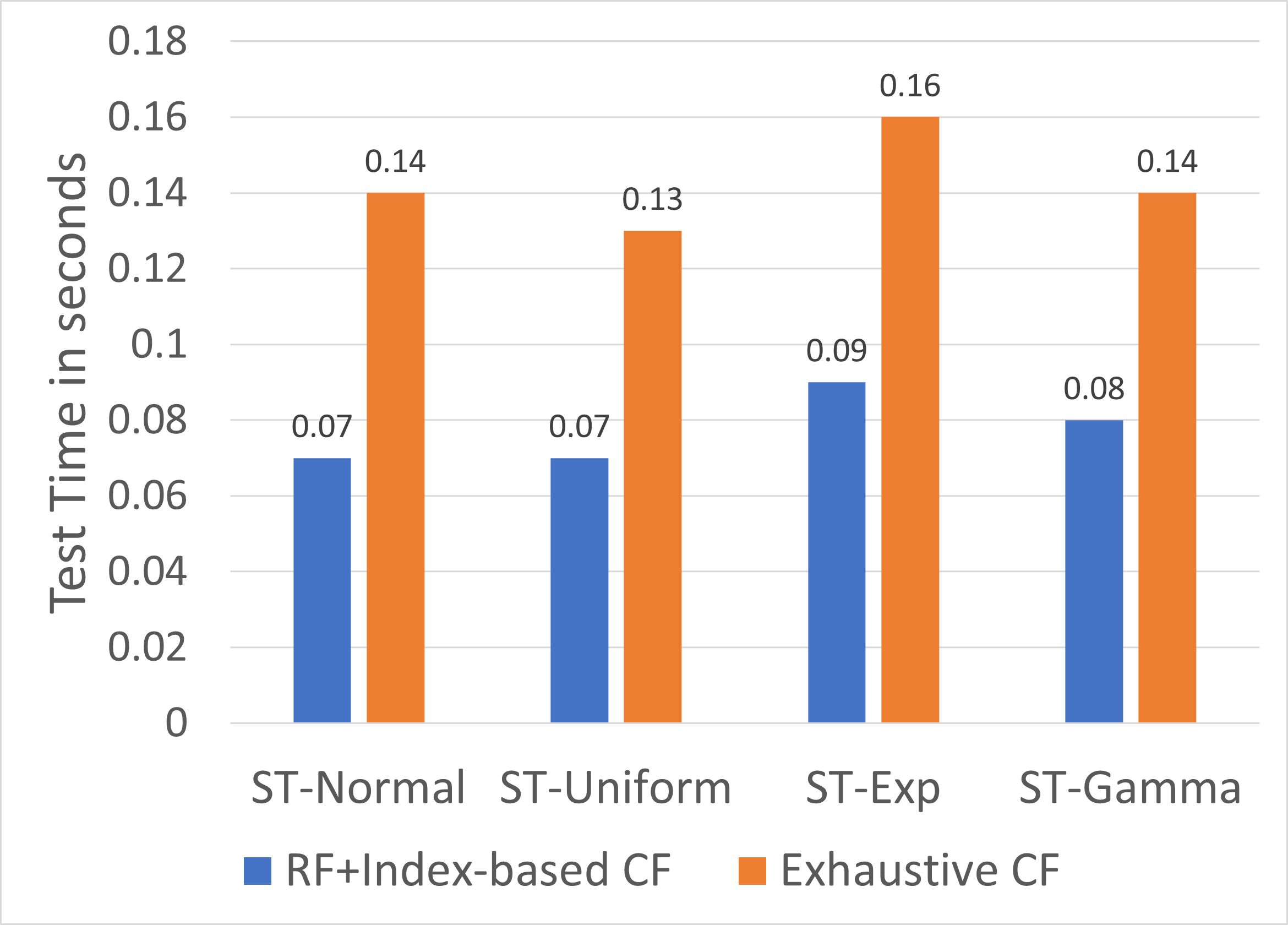}
		\caption{Prediction latency}
		\label{fig:CFTestTimeMGSynth}
	\end{subfigure}
	\begin{subfigure}[t]{0.235\textwidth}
		\centering
		\includegraphics[width=\linewidth]{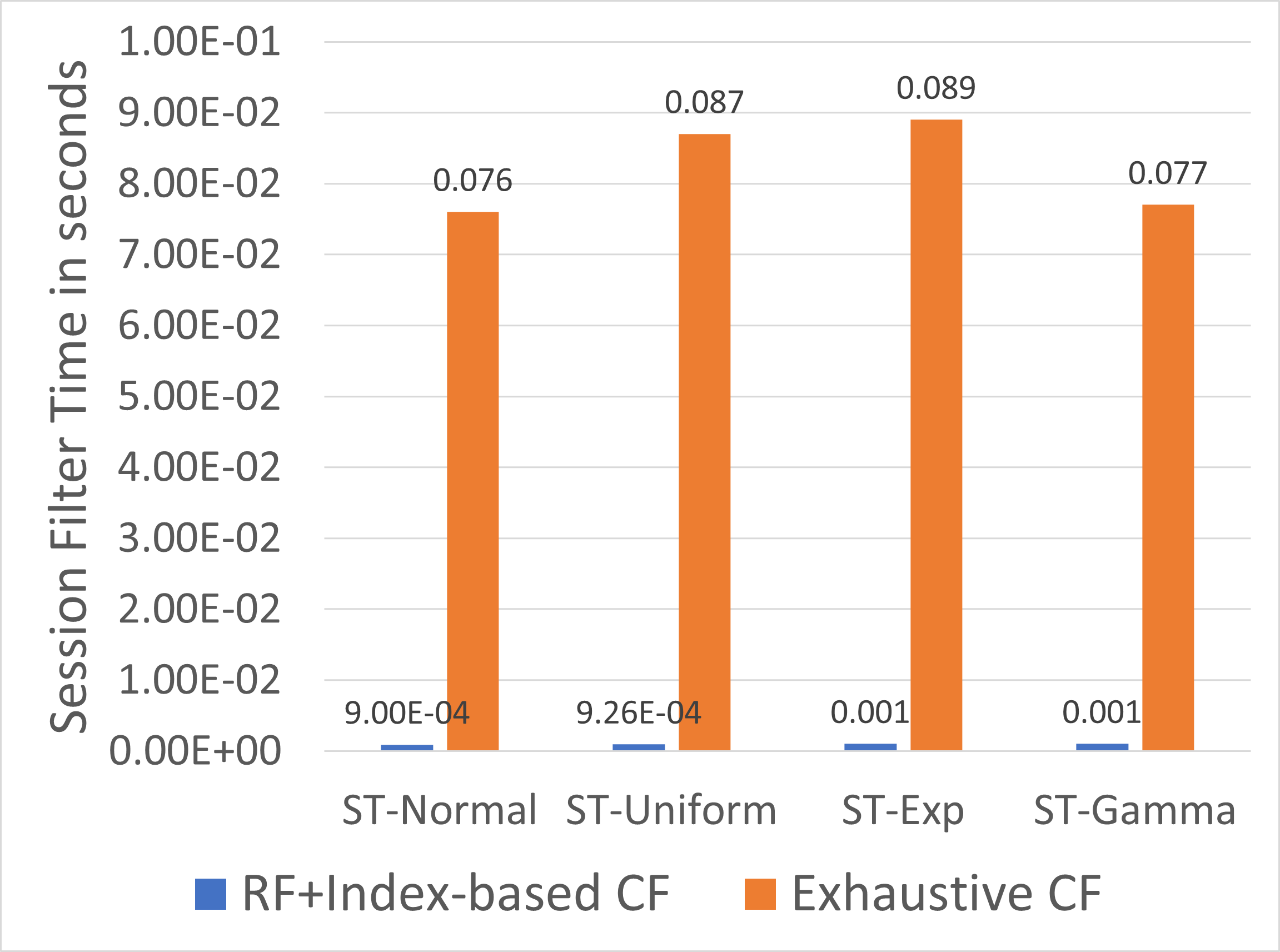}
		\caption{Session filtering latency}
		\label{fig:CFSessTimeMGSynth}
	\end{subfigure}
	\begin{subfigure}[t]{0.26\textwidth}
		\centering
		\includegraphics[width=\linewidth]{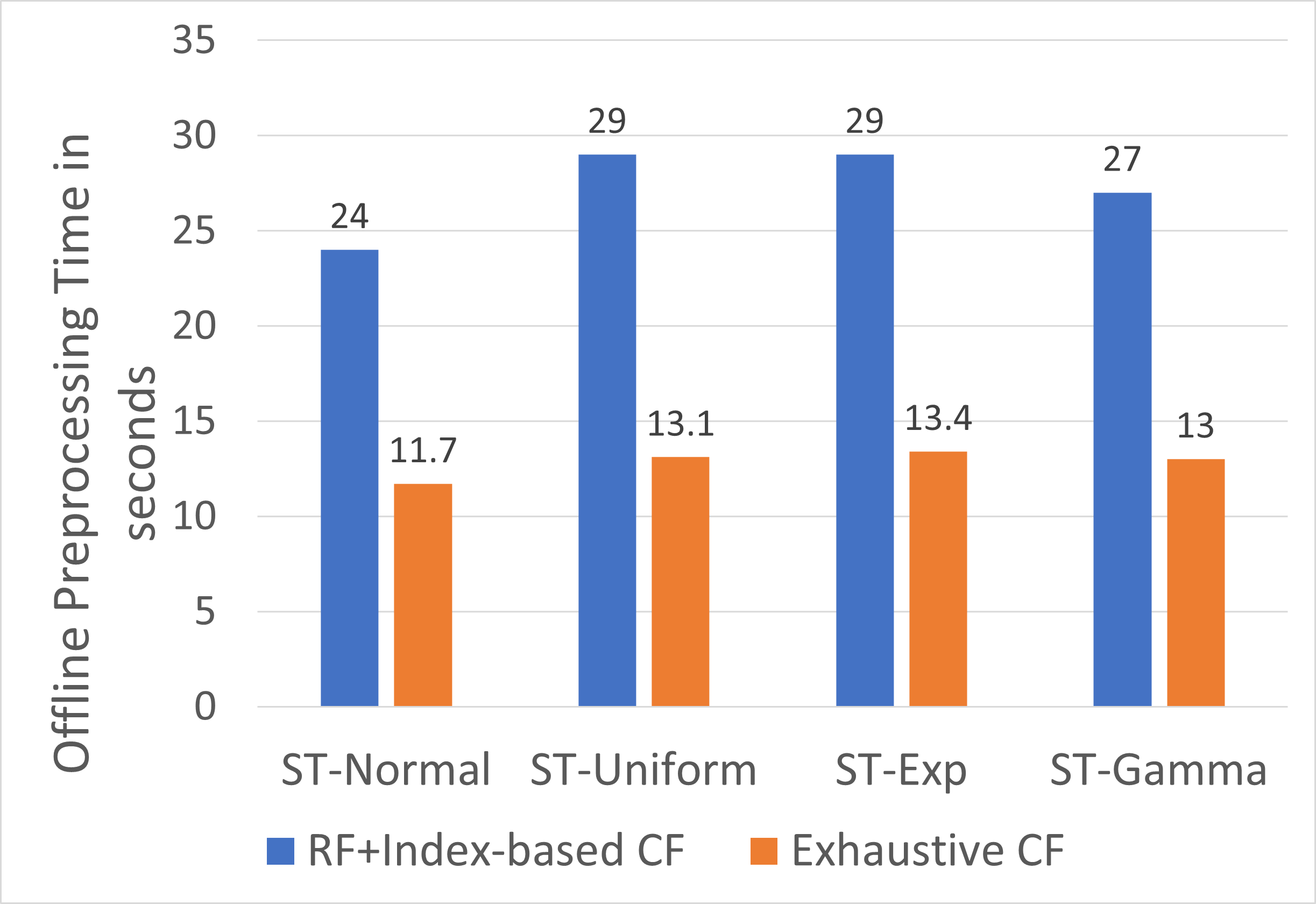}
		\caption{Pre-processing time}
		\label{fig:CFTrainTimeMGSynth}
	\end{subfigure}
	\caption{Comparing our two-step BI pattern prediction against the exhaustive CF baseline on the AHI dataset.}
	\label{fig:CFBaseline}
\end{figure*}

\begin{figure*}[tb]
	\centering
	\begin{subfigure}[t]{0.25\textwidth}
		\centering
		\includegraphics[width=\linewidth]{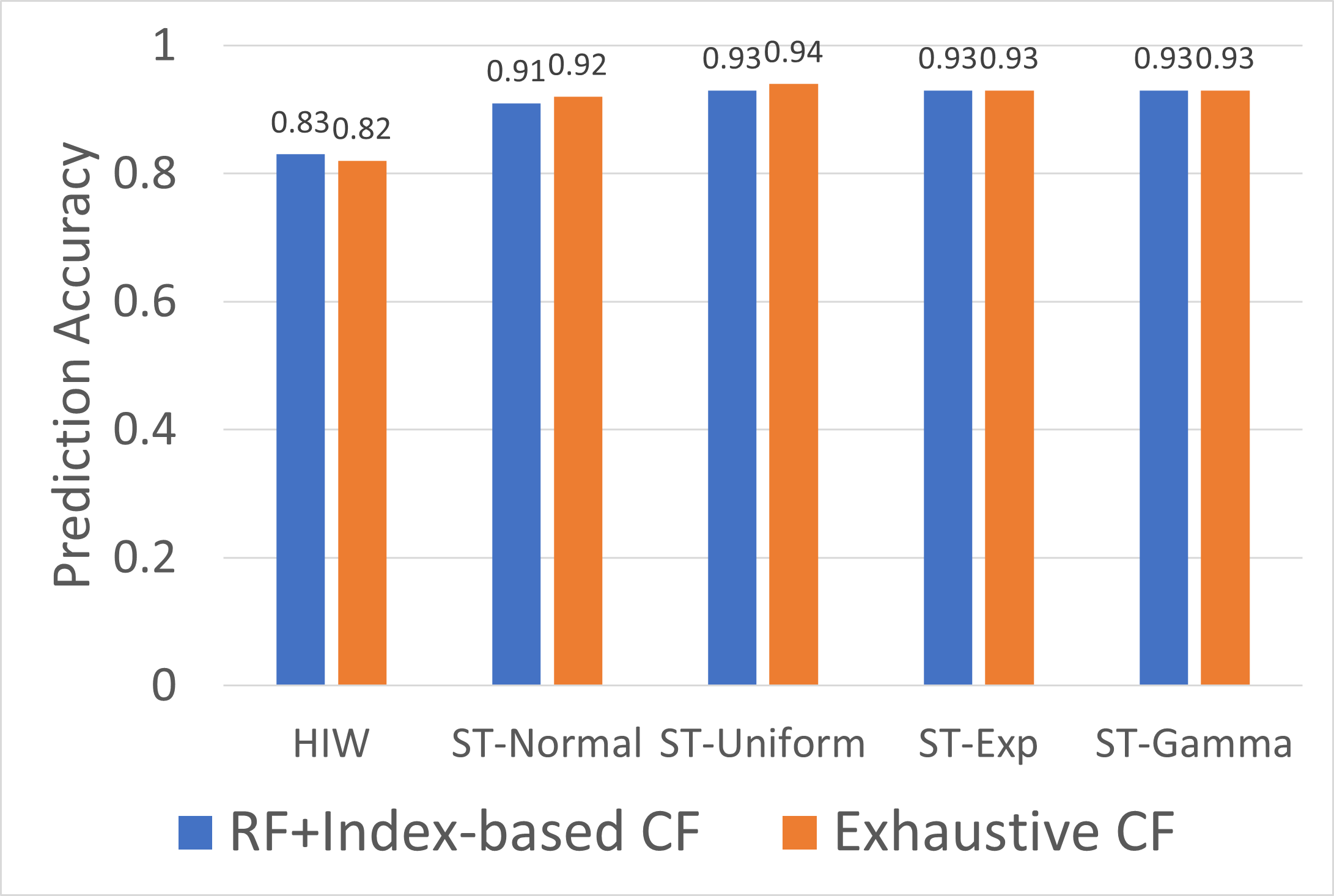}
		\caption{Prediction accuracy}
		\label{fig:CFQualityOrig}
	\end{subfigure}
	\begin{subfigure}[t]{0.23\textwidth}
		\centering
		\includegraphics[width=\linewidth]{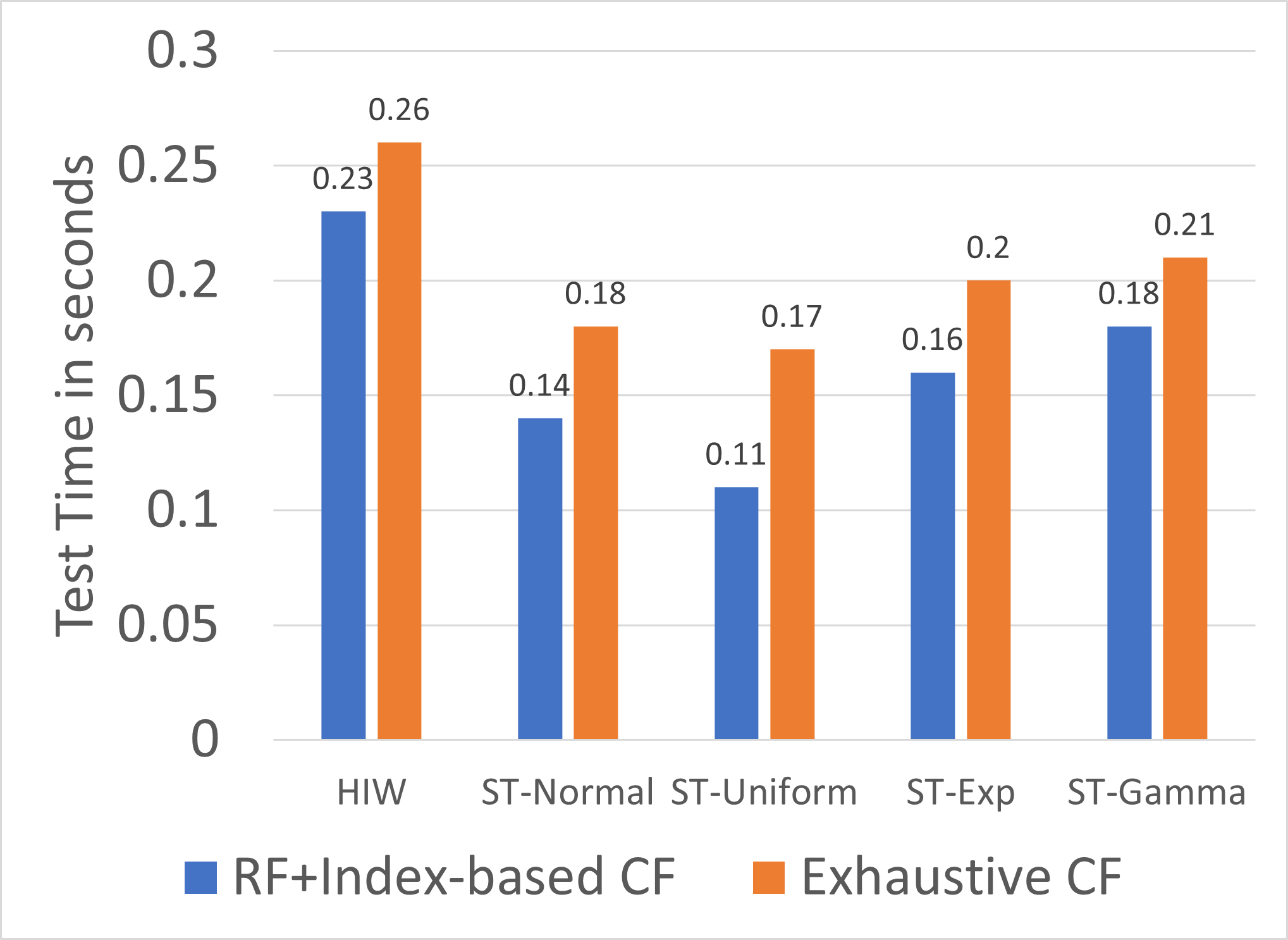}
		\caption{Prediction latency}
		\label{fig:CFTestTimeOrig}
	\end{subfigure}
	\begin{subfigure}[t]{0.25\textwidth}
		\centering
		\includegraphics[width=\linewidth]{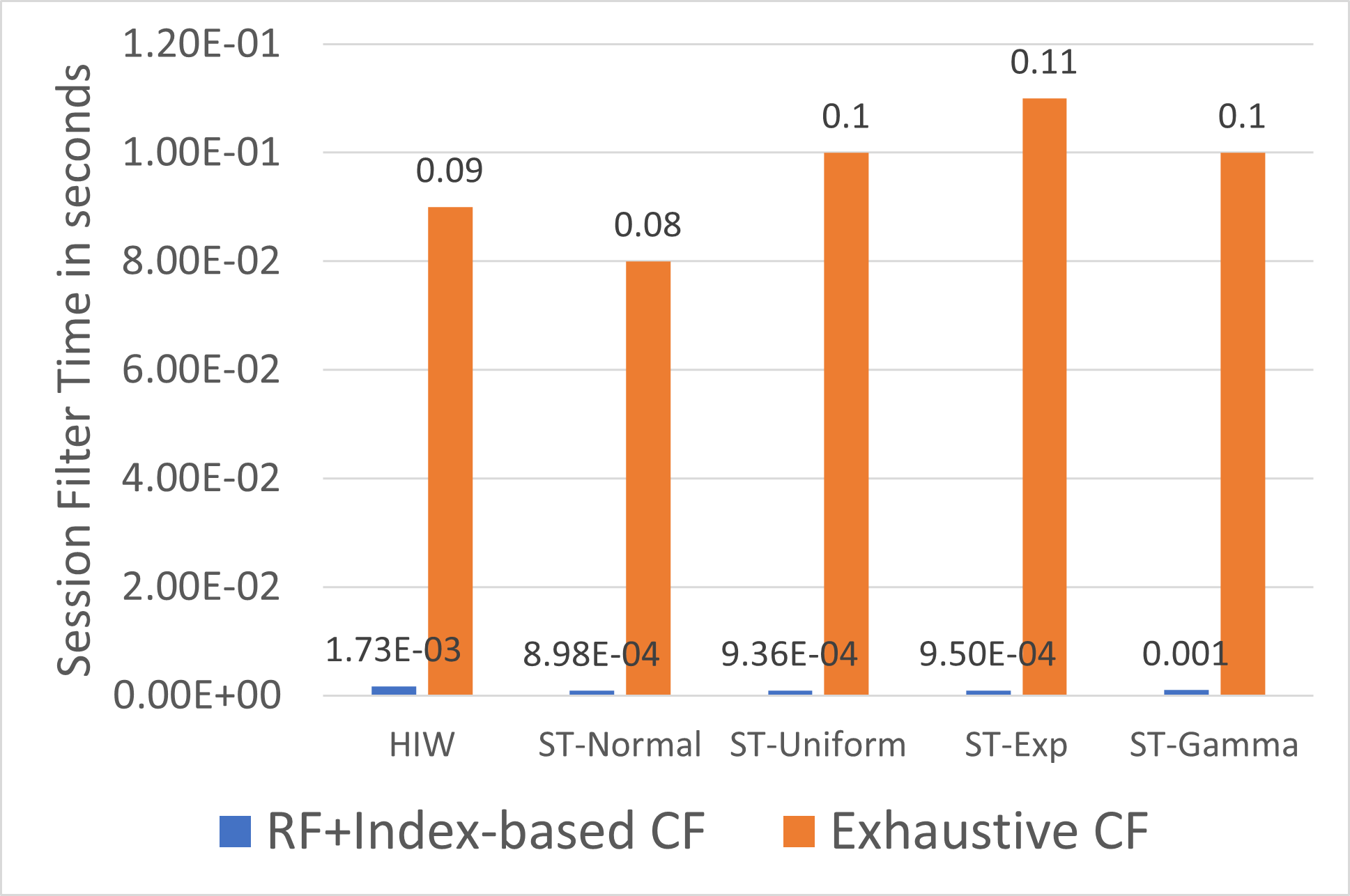}
		\caption{Session filtering latency}
		\label{fig:CFSessTimeOrig}
	\end{subfigure}
	\begin{subfigure}[t]{0.25\textwidth}
		\centering
		\includegraphics[width=\linewidth]{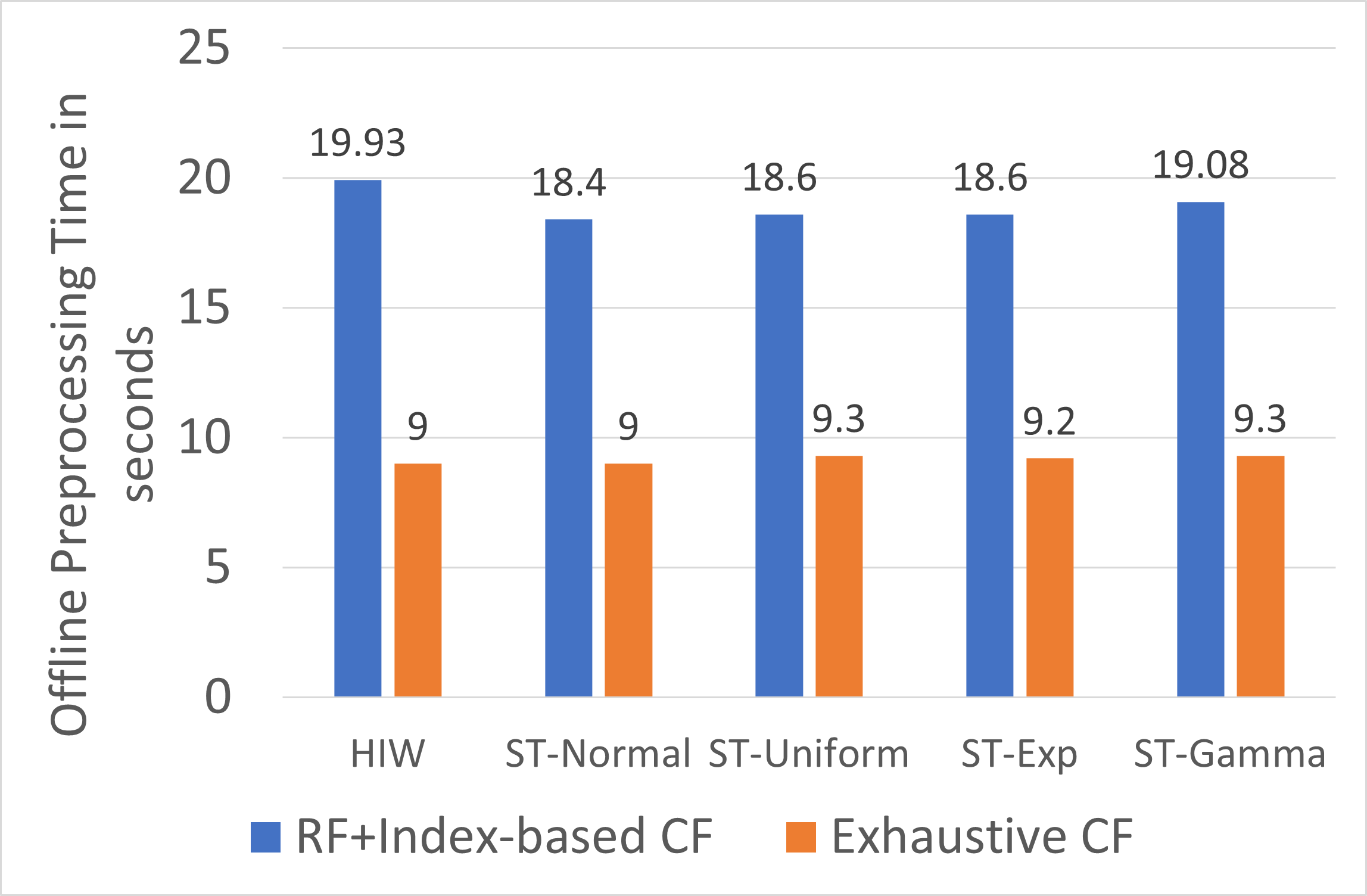}
		\caption{Pre-processing time}
		\label{fig:CFTrainTimeOrig}
	\end{subfigure}
	\caption{Comparing our two-Step BI pattern prediction against the exhaustive CF baseline on the HI dataset.}
	\label{fig:CFBaselineOrig}
\end{figure*}

Figures~\ref{fig:CFBaseline} and~\ref{fig:CFBaselineOrig} show a detailed comparison of our two-step approach with the exhaustive CF baseline on the AHI and HI datasets, respectively. Figures~\ref{fig:CFQualityMGSynth} and~\ref{fig:CFQualityOrig} show that the prediction accuracy with respect to top-$k$ BI patterns for HIW as well as various synthetic workloads for \sys~is comparable to that of the exhaustive CF baseline. Figures~\ref{fig:CFTestTimeMGSynth} and~\ref{fig:CFTestTimeOrig} compare the top-$k$ BI pattern prediction latency of \sys~with the exhaustive baseline. We see that \sys~outperforms the exhaustive baseline for both the datasets, while approximately achieving a 2$\times$ speedup for the AHI dataset. The reason for this impressive empirical result on the AHI dataset as compared to the HI dataset, is that there are fewer sessions per MG in the former compared to the latter (see Table~\ref{tab:workloads_user_session}). The two-step approach is thus able to exploit its pruning power to narrow down the number of relevant user sessions for making $P_{BI}$ recommendations.

Figures~\ref{fig:CFSessTimeMGSynth} and~\ref{fig:CFSessTimeOrig} compare the session filtering latency of the two approaches. We see that our $CF_{Index}$ approach provides a constant $O(1)$ access time latency as compared to the exhaustive CF baseline that requires comparing the ongoing test session with all the user sessions in the prior workload with an $O(|S| \cdot |V|)$ access time latency, where $|S|$ is the number of prior sessions and $|V|$ is the dimensionality of the session summary vectors. These experiments validate the effectiveness of our two step-approach for top-$k$ query prediction in pruning the search space to reduce prediction latency while maintaining comparable accuracy. 

Finally, Figures~\ref{fig:CFTrainTimeMGSynth} and~\ref{fig:CFTrainTimeOrig} compare the pre-processing time for the two approaches. Pre-processing time for \sys~includes the training time for the RF model, as well as the CF index construction time. For the exhaustive approach, this includes the session summary computation time. We observe that the pre-processing time of \sys~is much higher than that of the exhaustive CF baseline. However, since this is an offline process, the overhead is justified in order to get 
lower query prediction latencies at runtime.

\subsection{User Study}
\label{sec:user_study}

We conducted a detailed user study on the prototype implementation of \sys~against the HI dataset with 15 real-world users, including data scientists, data analysts and non-technical business users, to ascertain the quality and usefulness of the recommendations provided. 
The user study comprises of different session tasks containing MGs such as utilization of healthcare resources (UTILIZATION), costs covered by insurance (ALLOWED AMOUNT), net payments made by insurance (NET PAYMENT), etc. Each such session task is associated with a user session, wherein at each state in the session, the user issues a query to explore information about the task and the system provides a response to the query along with its top-$k$ $P_{BI}$ recommendations for the next possible state. 

Our user study contains three session tasks with one MG per session task. For each user query in a session, the participants were requested to select all the recommendations amongst the top-3 system recommendations that the user felt were interesting and useful with respect to the given user query. The users could also choose a ``none of the above'' option, if none of the recommendations seemed useful. We evaluate the quality of \sys~recommendations in terms of two metrics: (1) Precision@3, which is the percentage of total user responses where the user chose at least one of the top-3 system recommendations as useful, and 
(2) Mean Reciprocal Rank (MRR) (Equation~\ref{eqn:mrr}), where $rank_i$ is the ranked position of the system recommendation that received the most user votes, among the top-3 recommendations. For example, if \sys's second recommendation was the one that received the most user votes for a query $q_i$, then $rank_i = 2$. 
We compute MRR per session (or session task) by averaging the reciprocal ranks of the most voted system recommendations for each query, across all queries $Q$ in a session (or session task). 
\begin{equation}
    MRR = {1 \over{|Q|}} \sum^{|Q|}_{i=1}{1 \over rank_i}
    \label{eqn:mrr}
\end{equation}

Table~\ref{tab:user_study_results} shows the results of the user study in terms of Precision@3 and MRR for the three session tasks. 
We see that for all three tasks Precision@3 is high, with an average of 91.90\%. 
The user study results thus validates the effectiveness of \sys~in terms of making good quality recommendations that are useful to the users for guided data analysis for BI applications.

\sys~does reasonably well in terms of MRR with an average MRR of 0.62 across all three session tasks. We notice that the MRR score for $Task_{US_2}$ is lower compared to the other two tasks. Upon further investigation of $Task_{US_2}$ results, we saw that \sys~prioritized recommending BI patterns with $op_{BI}$ such as a PIVOT (switching the dimensions by which the analyzed measure was being sliced/diced by) over other BI patterns with $op_{BI}$ such as COMPARE (comparing two measures along a dimension) or TREND (analyzing the variation of measures over time). \sys~makes these BI pattern recommendations based on patterns learned from prior user sessions. However, some users in the study found the recommendations with the COMPARE and TREND operations more useful. We leave further investigation and improvement of BI recommendation rankings based on user feedback/preference as future work.
%We therefore see that there is a room for improvement in terms of the ranking of recommendations provided by \sys~that could be done based on current user feedback/preference. We leave the further refinement and fine tuning of our online $P_{BI}$ prediction models using feedback/preference from current users as future work. 

\begin{table}[h]
    \centering
    \caption{User study results.}
    \begin{tabular}{|c||c|c|c|}
        \hline
         & \textbf{$Task_{US_1}$} & \textbf{$Task_{US_2}$} & \textbf{$Task_{US_3}$}  \\ \hline
       Precision@3  & 88.9\% & 97.93\% & 88.9\% \\ \hline
       MRR  & 0.72 & 0.46 & 0.69 \\ \hline
    \end{tabular}
    \label{tab:user_study_results}
\end{table}

%% file: related.tex
%!TEX root = main.tex
\section{Related Work}
\label{sec:relatedWork}

\subsection{Recommendation Systems for Data Analysis.}

Recommendation systems for data analysis and exploration is an active area of research.
Most of these systems~\cite{DBLP:journals/dss/AligonGGMR15,10.1145/1651291.1651306,DBLP:conf/dawak/GiacomettiMN09} adopt a collaborative filtering (CF) approach for recommending OLAP sessions. Given an active user session, they use CF to search for relevant sessions in the logs of prior user sessions using similarity metrics described in~\cite{DBLP:journals/kais/AligonGMRT14} and recommend the top ranked sessions.  
Unlike our index-based CF approach, these proposed approaches are exhaustive in nature and would typically require for each user query to scan all prior sessions to find relevant sessions that match the active query sequence so far to make a recommendation.

Recommendation systems for interactive data analysis platforms such as IBM's Cognos Assistant~\cite{CognosAssistant}, Microsoft's PowerBI~\cite{PowerBi}, Kibana~\cite{Kibana}, Tableau~\cite{AskData}, and REACT~\cite{DBLP:conf/kdd/MiloS18} utilize past user interactions with the system to derive context which is used to generate next-action suggestions for users in similar contexts. Given a current user session and its associated context, these recommendation systems such as REACT find the top-$k$ similar sessions from prior query logs using k-Nearest Neighbors (kNN) search. 
\citet{DBLP:conf/dawak/AufaureKMRV13} 
learns a user behavior model by clustering queries from query logs, model user behavior as a Markov Model to predict the next query that the user is most likely to issue. Unlike \sys, these systems do not utilize the semantic information of the BI domain for reasoning about the analysis tasks to prune the search space of relevant sessions. Further, these systems being IDA platforms are often used as one shot Q\&A systems and consequently cannot exploit the conversational context across several data analysis steps to make recommendations relevant to the user's current state of data analysis.

~\citet{10.14778/2824032.2824103} propose a smart Drill-Down operator for data exploration systems that aims to discover interesting subsets of data with fewer operations. At each step the system recommends different dimensions to drill-down along to get to interesting insights based on a set of rules.
Auto-Suggest~\cite{DBLP:conf/sigmod/YanH20} learns analysis patterns in data science work flows from interactions of data scientists with Jupyter Notebooks. It suggests steps for data preparation in terms of the next operators while utilizing latent sequence correlation between operators and data characteristics. Unlike \sys, both systems limit the recommendations to single operators or BI operations such as Drill-Down, thereby limiting the guidance that they can provide for data exploration.

Another class of work in the area focuses on fully automating the process of exploratory data analysis using DL and reinforcement learning (RL) techniques~\cite{DBLP:conf/sigmod/MiloS20, DBLP:conf/sigmod/MiloS18, DBLP:conf/cikm/ElMS19}. The key idea is to use RL techniques to learn to perform EDA operations by independently interacting with the dataset and learning methods for predicting users' preferences. Our \sys\ on the other hand focuses on providing guidance at each step of an active data exploration session using models that learn from prior user experiences and semantic knowledge of the BI domain to provide relevant recommendations based on the current conversational context.

Active-learning approaches for recommendations for interactive data exploration systems~\cite{DBLP:journals/tkde/DimitriadouPD16, DBLP:journals/pvldb/HuangPPALD18} are based on the principle of {\em explore-by-example} wherein the system collects user preferences on samples of data or objects in terms of relevance to the desired analysis task, to build a user profile/interest. This user input is used to train a classification model to classify database objects as relevant or irrelevant to the user's analysis task and refine the data samples iteratively to guide user through the process exploring the dataset. These systems typically do not exploit past user experiences and do not actually suggest BI queries and operations as suggested next-steps in data analysis. 

\subsection{Conversational Recommendation Systems}

Conversational recommendation systems have seen a lot of interest in the recent past. These include systems~\cite{NEURIPS2018_800de15c, 10.1145/2939672.2939746, 10.1145/3336191.3371769} that typically employ <user-item> based CF approaches while incorporating the conversational context between the system and the user to recommend items such as movies or restaurants based on the ratings provided by users on the items. 
%R. Li et al.~
\citet{NEURIPS2018_800de15c} use a deep learning model (hierarchical auto-encoder) to capture the conversational context which is trained on a set of conversations centered around the theme of providing movie recommendations. 
\citet{10.1145/2939672.2939746} focus on the problem of cold start in conversational recommendation systems in cases where users have no history of rating items. The new user's profile is built by asking clarifying questions generated using active learning coupled with multi-armed bandit models. Those models balance the explore and exploit paradigms by minimizing the number of questions asked while maximizing the information gain.

Conversational recommendation systems that utilize user feedback to fine tune their prediction models include~\cite{10.1145/3336191.3371769, 10.1145/1557914.1557930}. 
\citet{10.1145/3336191.3371769} train a matrix factorization model on users and items, which is updated based on the positive and negative feedback provided by users to provide refined recommendations. 
\citet{10.1145/1557914.1557930} use Markov Decision Models for making recommendations. The actions are adaptively updated based on user feedback using a reinforcement learning model that chooses actions based on the type of user (novice or experienced) and rewards based on whether a user browses the recommendations or decides to add them to their travel plan.

In general, these systems do not directly address the problem of guided data analysis for BI. However, some of the techniques introduced in these systems, such as capturing the conversational context, address the cold start problem, and updating the prediction models based on user feedback, are complementary to our work of building systems that provide guidance for data analysis.

%% file: conclusion.tex
%!TEX root = main.tex
%\vspace{-5pt}
\section{Conclusions}
\label{sec:conclusion}

In this work, we proposed a system, \sys, for guided conversational analysis. We model an analysis state as a graph, combining information from prior workloads and semantic information from the BI ontology. We use GraphSAGE~\cite{NIPS2017_5dd9db5e} for representation learning of the state graphs, allowing us to work with compact vectors in our recommendation algorithms. 
Our proposed two-step approach for BI pattern prediction is extremely effective in pruning the search space of relevant sessions which makes model training for high-level $BI_{Intent}$ prediction simpler and more effective. Simple multi-class classifiers have been shown to achieve high predictive accuracy with less training data. As demonstrated by our experimental evaluation and user study, our BI query prediction using prior user experience, and semantic knowledge from the BI ontology provides useful recommendations 
with low latency making them suitable for interactive conversational systems. 